\documentclass{article}

\usepackage{pgfplots}
\pgfplotsset{compat=1.7}

\usepackage{graphicx} 
\usepackage[utf8]{inputenc}
\usepackage[colorlinks=true]{hyperref}

\usepackage[margin=1in]{geometry}
\usepackage{pgfmath,pgffor}
\usepackage[tight,footnotesize]{subfigure}
\usepackage{epstopdf}
\graphicspath{{./Figures/}}
\usepackage{upgreek}
\usepackage{amsmath,amssymb,amsthm}

\usepackage{xcolor}
\definecolor{MyGrey5}{RGB}{220,220,220}
\definecolor{MyGrey}{RGB}{234,234,234}
\definecolor{MyBlue}{RGB}{0,0,255}
\definecolor{myLightSlateGray}{rgb}{0.46484,0.53125,0.59766}
\definecolor{myGray}{rgb}{0.5,0.5,0.5}
\definecolor{myFireBrick}{rgb}{ 0.6953   ,  0.1328  ,  0.1328}

\definecolor{myBlue}{RGB}{0,0,255}
\definecolor{myGreen}{rgb}{ 0          ,  0.5000  ,  0     }
\definecolor{myPurple}{rgb}{0.5     ,    0  ,0.5}
\definecolor{myRed}{rgb}{ 1        ,  0.0000  ,  0     }
\definecolor{myDarkSlateGray}{rgb}{0.1836,  0.3086  ,  0.3086}
\definecolor{myChocolate}{rgb}{0.8203 ,  0.4102  ,  0.1172}
\definecolor{myFireBrick}{rgb}{ 0.6953   ,  0.1328  ,  0.1328}
\definecolor{myGrey}{rgb}{0.5,0.5,0.5}
\definecolor{myOrange}{rgb}{1,0.64453,0}
\definecolor{myCyan}{rgb}{0,1,1}
\definecolor{WatermarkGrey}{rgb}{0.75,0.75,0.75}

\definecolor{matlabBlue}{rgb}{   0 ,   0.4470,    0.7410}
\definecolor{matlabOrange}{rgb}{0.8500 ,   0.3250,    0.0980}

\definecolor{myWhite}{rgb}{1,1,1}
\definecolor{mySnow}{rgb}{1,0.97917,0.97917}
\definecolor{myHoneydew}{rgb}{0.9375,1,0.9375}
\definecolor{myMintCream}{rgb}{0.95833,1,0.97917}
\definecolor{myAzure}{rgb}{0.9375,1,1}
\definecolor{myAliceBlue}{rgb}{0.9375,0.97083,1}
\definecolor{myGhostWhite}{rgb}{0.97083,0.97083,1}
\definecolor{myWhiteSmoke}{rgb}{0.95833,0.95833,0.95833}
\definecolor{mySeashell}{rgb}{1,0.95833,0.92969}
\definecolor{myBeige}{rgb}{0.95833,0.95833,0.85938}
\definecolor{myOldLace}{rgb}{0.99167,0.95833,0.89844}
\definecolor{myFloralWhite}{rgb}{1,0.97917,0.9375}
\definecolor{myIvory}{rgb}{1,1,0.9375}
\definecolor{myAntiqueWhite}{rgb}{0.97917,0.91797,0.83984}
\definecolor{myLinen}{rgb}{0.97917,0.9375,0.89844}
\definecolor{myLavenderBlush}{rgb}{1,0.9375,0.95833}
\definecolor{myMistyRose}{rgb}{1,0.89063,0.87891}
\definecolor{myGray}{rgb}{0.5,0.5,0.5}
\definecolor{myGainsboro}{rgb}{0.85938,0.85938,0.85938}
\definecolor{myLightGray}{rgb}{0.82422,0.82422,0.82422}
\definecolor{mySilver}{rgb}{0.75,0.75,0.75}
\definecolor{myDarkGray}{rgb}{0.66016,0.66016,0.66016}
\definecolor{myDimGray}{rgb}{0.41016,0.41016,0.41016}
\definecolor{myLightSlateGray}{rgb}{0.46484,0.53125,0.59766}
\definecolor{mySlateGray}{rgb}{0.4375,0.5,0.5625}
\definecolor{myDarkSlateGray}{rgb}{0.18359,0.30859,0.30859}
\definecolor{myBlack}{rgb}{0,0,0}
\definecolor{myRed}{rgb}{1,0,0}
\definecolor{myLightSalmon}{rgb}{1,0.625,0.47656}
\definecolor{mySalmon}{rgb}{0.97917,0.5,0.44531}
\definecolor{myDarkSalmon}{rgb}{0.91016,0.58594,0.47656}
\definecolor{myLightCoral}{rgb}{0.9375,0.5,0.5}
\definecolor{myIndianRed}{rgb}{0.80078,0.35938,0.35938}
\definecolor{myCrimson}{rgb}{0.85938,0.078125,0.23438}
\definecolor{myFireBrick}{rgb}{0.69531,0.13281,0.13281}
\definecolor{myDarkRed}{rgb}{0.54297,0,0}
\definecolor{myPink}{rgb}{1,0.75,0.79297}
\definecolor{myLightPink}{rgb}{1,0.71094,0.75391}
\definecolor{myHotPink}{rgb}{1,0.41016,0.70313}
\definecolor{myDeepPink}{rgb}{1,0.078125,0.57422}
\definecolor{myPaleVioletRed}{rgb}{0.85547,0.4375,0.57422}
\definecolor{myMediumVioletRed}{rgb}{0.77734,0.082031,0.51953}
\definecolor{myOrange}{rgb}{1,0.64453,0}
\definecolor{myDarkOrange}{rgb}{1,0.54688,0}
\definecolor{myCoral}{rgb}{1,0.49609,0.3125}
\definecolor{myTomato}{rgb}{1,0.38672,0.27734}
\definecolor{myOrangeRed}{rgb}{1,0.26953,0}
\definecolor{myYellow}{rgb}{1,1,0}
\definecolor{myLightYellow}{rgb}{1,1,0.875}
\definecolor{myLemonChiffon}{rgb}{1,0.97917,0.80078}
\definecolor{myLightGoldenrodYellow}{rgb}{0.97917,0.97917,0.82031}
\definecolor{myPapayaWhip}{rgb}{1,0.93359,0.83203}
\definecolor{myMoccasin}{rgb}{1,0.89063,0.70703}
\definecolor{myPeachPuff}{rgb}{1,0.85156,0.72266}
\definecolor{myPaleGoldenrod}{rgb}{0.92969,0.90625,0.66406}
\definecolor{myKhaki}{rgb}{0.9375,0.89844,0.54688}
\definecolor{myDarkKhaki}{rgb}{0.73828,0.71484,0.41797}
\definecolor{myGold}{rgb}{1,0.83984,0}
\definecolor{myBrown}{rgb}{0.64453,0.16406,0.16406}
\definecolor{myCornsilk}{rgb}{1,0.97083,0.85938}
\definecolor{myBlanchedAlmond}{rgb}{1,0.91797,0.80078}
\definecolor{myBisque}{rgb}{1,0.89063,0.76563}
\definecolor{myNavajoWhite}{rgb}{1,0.86719,0.67578}
\definecolor{myWheat}{rgb}{0.95833,0.86719,0.69922}
\definecolor{myBurlyWood}{rgb}{0.86719,0.71875,0.52734}
\definecolor{myTan}{rgb}{0.82031,0.70313,0.54688}
\definecolor{myRosyBrown}{rgb}{0.73438,0.55859,0.55859}
\definecolor{mySandyBrown}{rgb}{0.95417,0.64063,0.375}
\definecolor{myGoldenrod}{rgb}{0.85156,0.64453,0.125}
\definecolor{myDarkGoldenrod}{rgb}{0.71875,0.52344,0.042969}
\definecolor{myPeru}{rgb}{0.80078,0.51953,0.24609}
\definecolor{myChocolate}{rgb}{0.82031,0.41016,0.11719}
\definecolor{mySaddleBrown}{rgb}{0.54297,0.26953,0.074219}
\definecolor{mySienna}{rgb}{0.625,0.32031,0.17578}
\definecolor{myMaroon}{rgb}{0.5,0,0}
\definecolor{myGreen}{rgb}{0,0.5,0}
\definecolor{myPaleGreen}{rgb}{0.59375,0.98333,0.59375}
\definecolor{myLightGreen}{rgb}{0.5625,0.92969,0.5625}
\definecolor{myYellowGreen}{rgb}{0.60156,0.80078,0.19531}
\definecolor{myGreenYellow}{rgb}{0.67578,1,0.18359}
\definecolor{myChartreuse}{rgb}{0.49609,1,0}
\definecolor{myLawnGreen}{rgb}{0.48438,0.9875,0}
\definecolor{myLime}{rgb}{0,1,0}
\definecolor{myLimeGreen}{rgb}{0.19531,0.80078,0.19531}
\definecolor{myMediumSpringGreen}{rgb}{0,0.97917,0.60156}
\definecolor{mySpringGreen}{rgb}{0,1,0.49609}
\definecolor{myMediumAquamarine}{rgb}{0.39844,0.80078,0.66406}
\definecolor{myAquamarine}{rgb}{0.49609,1,0.82813}
\definecolor{myLightSeaGreen}{rgb}{0.125,0.69531,0.66406}
\definecolor{myMediumSeaGreen}{rgb}{0.23438,0.69922,0.44141}
\definecolor{mySeaGreen}{rgb}{0.17969,0.54297,0.33984}
\definecolor{myDarkSeaGreen}{rgb}{0.55859,0.73438,0.55859}
\definecolor{myForestGreen}{rgb}{0.13281,0.54297,0.13281}
\definecolor{myDarkGreen}{rgb}{0,0.39063,0}
\definecolor{myOliveDrab}{rgb}{0.41797,0.55469,0.13672}
\definecolor{myOlive}{rgb}{0.5,0.5,0}
\definecolor{myDarkOliveGreen}{rgb}{0.33203,0.41797,0.18359}
\definecolor{myTeal}{rgb}{0,0.5,0.5}
\definecolor{myBlue}{rgb}{0,0,1}
\definecolor{myLightBlue}{rgb}{0.67578,0.84375,0.89844}
\definecolor{myPowderBlue}{rgb}{0.6875,0.875,0.89844}
\definecolor{myPaleTurquoise}{rgb}{0.68359,0.92969,0.92969}
\definecolor{myTurquoise}{rgb}{0.25,0.875,0.8125}
\definecolor{myMediumTurquoise}{rgb}{0.28125,0.81641,0.79688}
\definecolor{myDarkTurquoise}{rgb}{0,0.80469,0.81641}
\definecolor{myLightCyan}{rgb}{0.875,1,1}
\definecolor{myCyan}{rgb}{0,1,1}
\definecolor{myAqua}{rgb}{0,1,1}
\definecolor{myDarkCyan}{rgb}{0,0.54297,0.54297}
\definecolor{myCadetBlue}{rgb}{0.37109,0.61719,0.625}
\definecolor{myLightSteelBlue}{rgb}{0.6875,0.76563,0.86719}
\definecolor{mySteelBlue}{rgb}{0.27344,0.50781,0.70313}
\definecolor{myLightSkyBlue}{rgb}{0.52734,0.80469,0.97917}
\definecolor{mySkyBlue}{rgb}{0.52734,0.80469,0.91797}
\definecolor{myDeepSkyBlue}{rgb}{0,0.74609,1}
\definecolor{myDodgerBlue}{rgb}{0.11719,0.5625,1}
\definecolor{myCornflowerBlue}{rgb}{0.39063,0.58203,0.92578}
\definecolor{myRoyalBlue}{rgb}{0.25391,0.41016,0.87891}
\definecolor{myMediumBlue}{rgb}{0,0,0.80078}
\definecolor{myDarkBlue}{rgb}{0,0,0.54297}
\definecolor{myNavy}{rgb}{0,0,0.5}
\definecolor{myMidnightBlue}{rgb}{0.097656,0.097656,0.4375}
\definecolor{myPurple}{rgb}{0.5,0,0.5}
\definecolor{myLavender}{rgb}{0.89844,0.89844,0.97917}
\definecolor{myThistle}{rgb}{0.84375,0.74609,0.84375}
\definecolor{myPlum}{rgb}{0.86328,0.625,0.86328}
\definecolor{myViolet}{rgb}{0.92969,0.50781,0.92969}
\definecolor{myOrchid}{rgb}{0.85156,0.4375,0.83594}
\definecolor{myFuchsia}{rgb}{1,0,1}
\definecolor{myMagenta}{rgb}{1,0,1}
\definecolor{myMediumOrchid}{rgb}{0.72656,0.33203,0.82422}
\definecolor{myMediumPurple}{rgb}{0.57422,0.4375,0.85547}
\definecolor{myAmethyst}{rgb}{0.59766,0.39844,0.79688}
\definecolor{myBlueViolet}{rgb}{0.53906,0.16797,0.88281}
\definecolor{myDarkViolet}{rgb}{0.57813,0,0.82422}
\definecolor{myDarkOrchid}{rgb}{0.59766,0.19531,0.79688}
\definecolor{myDarkMagenta}{rgb}{0.54297,0,0.54297}
\definecolor{mySlateBlue}{rgb}{0.41406,0.35156,0.80078}
\definecolor{myDarkSlateBlue}{rgb}{0.28125,0.23828,0.54297}
\definecolor{myMediumSlateBlue}{rgb}{0.48047,0.40625,0.92969}
\definecolor{myIndigo}{rgb}{0.29297,0,0.50781}
\definecolor{myGrey}{rgb}{0.5,0.5,0.5}
\definecolor{myLightGrey}{rgb}{0.82422,0.82422,0.82422}
\definecolor{myDarkGrey}{rgb}{0.66016,0.66016,0.66016}
\definecolor{myDimGrey}{rgb}{0.41016,0.41016,0.41016}
\definecolor{myLightSlateGrey}{rgb}{0.46484,0.53125,0.59766}
\definecolor{mySlateGrey}{rgb}{0.4375,0.5,0.5625}
\definecolor{myDarkSlateGrey}{rgb}{0.18359,0.30859,0.30859}

\usepackage{multirow}
\usepackage{adjustbox}

\usepackage{cancel}
\usepackage{pgfplots}
\usepackage{pgfplotstable}

\newcommand\solidMrule[1][1cm]{\rule[0.5ex]{#1}{1.45pt}}

\newcommand\solidXXLrule[1][1cm]{\rule[0.001ex]{#1}{4pt}}

\newcommand\dashedruleM{\mbox{\solidMrule[2mm]\hspace{2mm}\solidMrule[2mm]\hspace{2mm}\solidMrule[2mm]}}

\newcommand\dotruleM{\mbox{\solidMrule[0.5mm]\hspace{0.5mm}\solidMrule[0.5mm]\hspace{0.5mm}\solidMrule[0.5mm]\hspace{0.5mm}\solidMrule[0.5mm]\hspace{0.5mm}\solidMrule[0.5mm]\hspace{0.5mm}\solidMrule[0.5mm]\hspace{0.5mm}\solidMrule[0.5mm]\hspace{0.5mm}\solidMrule[0.5mm]\hspace{0.5mm}\solidMrule[0.5mm]\hspace{0.5mm}\solidMrule[0.5mm]}}

\newcommand\dashdotruleM{\mbox{\solidMrule[2mm]\hspace{0.5mm}\solidMrule[1mm]\hspace{0.5mm}\solidMrule[2mm]\hspace{0.5mm}\solidMrule[1mm]\hspace{0.5mm}\solidMrule[2mm]}}

\title{Deep Learning Estimation of Modified Zernike Coefficients and Recovery of Point Spread Functions in Turbulence}

\author{Abu Bucker Siddik$^{*}$, Steven Sandoval, David Voelz, Laura E Boucheron, and Luis Varela}
\date{Klipsch School of Electrical and Computer Engineering, New Mexico State University, \\Las Cruces, New Mexico 88003, USA\\
$^*$siddik@nmsu.edu}

\begin{document}

\maketitle

\begin{abstract}
Recovering the turbulence-degraded point spread function from a single intensity image is important for a variety of imaging applications. Here, a deep learning model based on a convolutional neural network is applied to intensity images to predict a modified set of Zernike polynomial coefficients corresponding to wavefront aberrations in the pupil due to turbulence. The modified set assigns an absolute value to coefficients of even radial orders due to a sign ambiguity associated with this problem and is shown to be sufficient for specifying the intensity point spread function. Simulated image data of a point object and simple extended objects over a range of turbulence and detection noise levels are created for the learning model. The MSE results for the learning model show that the best prediction is found when observing a point object, but it is possible to recover a useful set of modified Zernike coefficients from an extended object image that is subject to detection noise and turbulence.    
   
\end{abstract}

\section{Introduction}
\label{sec:Introduction}
The point spread function (PSF) refers to the impulse response of an imaging system \cite{li2018fly}. When the PSF is known, it can be used for the correction of blur and other artifacts in images that are due to the system’s response. For example, in a space-invariant imaging situation, a PSF correction might be applied in a deconvolution step. Additionally, the propagation of light through a medium such as the atmosphere introduces wavefront aberrations at the aperture plane that further degrade the images. Therefore, the estimation of the combined system and medium PSF can potentially quantify the aberrations so that image correction can be performed.

\subsection{Imaging Model}
\label{subsection: Imaging Model}
Incoherent imaging of an object can be modeled using a linear space-invariant forward model
\begin{equation}
    I(x,y) = h(x,y) * I_0(x,y),
    \label{eq:I(x,y)}
\end{equation}
where $I(x,y)$ is the intensity image of a source object $I_0(x,y)$, $*$ is the convolution operator, and $h(x,y)$ is the intensity PSF given by 
\begin{equation}
    h(x,y)\propto \big| \mathcal{F}\big\{p(x,y)\mathrm{e}^{\,\mathrm{j}\upphi(x,y)}\big\} \big|^2,
    \label{eq:h(x,y)}
\end{equation}
where $\mathcal{F}$ is the Fourier transform, $|\cdot|$ is the modulus operator, $p(x,y)$ is the pupil function, and $\upphi(x,y)$ is the wavefront phase distortion applied at the pupil plane \cite{lu2022mitigating}. 

\subsection{Representing Wavefront Distortions}
\label{sec:RepresentingWavefrontDistortions}
\begin{figure*}[tbh!]
    \centering
    \includegraphics[width=\linewidth]{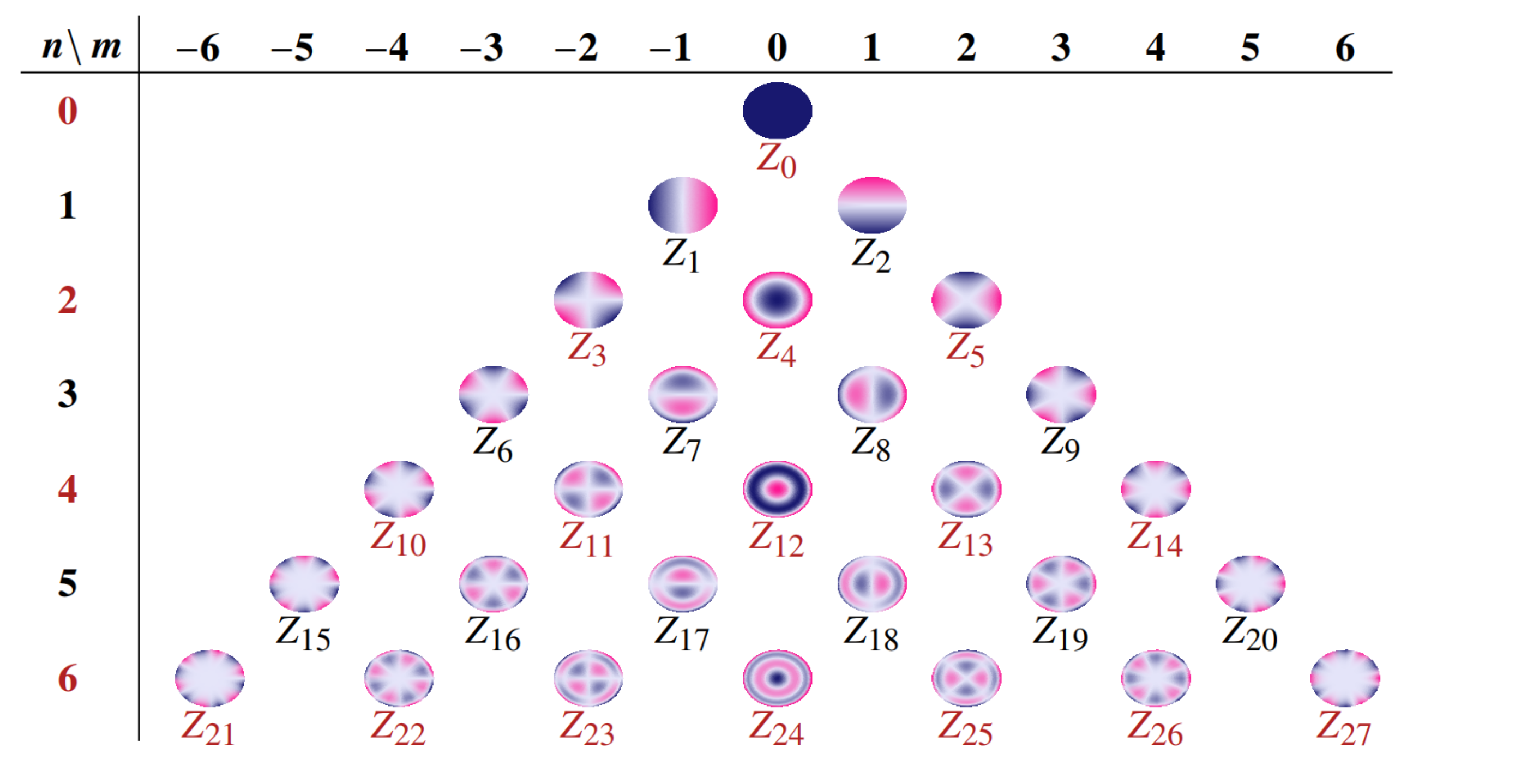}
    \caption{The first $28$ Zernike polynomials arranged in a pyramid structure. The rows follow radial order $n$ and columns follow angular frequency $m$ \cite{lakshminarayanan2011zernike}. The angularly even polynomials which are subject to ambiguity in intensity images correspond to even values of $n$ (corresponding labels are colored \textcolor{myFireBrick}{red}). In the colormap, positive values are indicated as \textcolor{myMidnightBlue}{\solidXXLrule[3.5mm]}, zero values are indicated as \textcolor{myLavender}{\solidXXLrule[3.5mm]}, and negative value are indicated as
\textcolor{myDeepPink}{\solidXXLrule[3.5mm]}.} 
    \label{tbl:ZernikeTable}
\end{figure*}

Although the PSF can be parameterized in various ways, to leverage results from disciplines such as adaptive optics, we consider the wavefront distortion $\upphi(x,y)$ related to the PSF through Eq.~(2). The phase distortion can be represented by a linear superposition of Zernike polynomials 
\begin{equation}
    \upphi(x,y) = \sum_{n,m} a_{nm}Z_n^m(x,y),
    \label{eq:upphi}
\end{equation}
where $Z_n^m$ are the Zernike polynomials and $a_{nm}$ are the Zernike coefficients \cite{janout2017psf, jing2010new, jin2018machine, schwiegerling2014optical}. Zernike polynomials form a set of orthogonal functions defined over the unit circle and may be expressed in a double $(m,n)$ indexing scheme \cite{lakshminarayanan2011zernike} as
\begin{equation}
Z_{n}^{m}(\rho ,\varphi )=\begin{cases}
			R_{n}^{\lvert m \rvert}(\rho)\cos(m\varphi ), & \text{if $m\geq 0$ }\\
            R_{n}^{\lvert m \rvert}(\rho)\sin(m\varphi ), & \text{if $m<0$ }
		 \end{cases}
\end{equation}
where $\rho$ is the normalized radial distance ($0\leq \rho \leq 1$), $\varphi$ is the azimuthal angle, $n$ is the radial order, $m$ is the angular frequency with $n \geq m \geq 0$, and $R_{n}^{\lvert m \rvert}(\rho )$ 
are radial polynomials given by 
\begin{equation}
R_{n}^{\lvert m \rvert}(\rho )=\begin{cases}
			\displaystyle\sum _{p=0}^{\tfrac {n-\lvert m \rvert}{2}}{\frac {(-1)^{p}\,(n-p)!}{p!\left({\tfrac {n+\lvert m \rvert}{2}}-p\right)!\left({\tfrac {n-\lvert m \rvert}{2}}-p\right)!}}\;\rho ^{n-2p}, & \text{if $n-\lvert m \rvert=0 \pmod 2$ }\\
            0, & \text{otherwise. }
		 \end{cases}
\end{equation}
 For our work, it is more convenient to utilize the single-indexed Zernike polynomials $Z_q$ where the index $q$ is related (OSA/ANSI standard) to the double-index by 
\begin{equation}
    q = \dfrac{n(n+2)+m}{2}.
\end{equation}
Zernike polynomials can be illustrated in a methodical way using a pyramid structure as given in Fig.~\ref{tbl:ZernikeTable}. In the figure, the indexing starts with $q=0$ on the top, then moves down the pyramid while scanning left to right. The rows of the pyramid correspond to the radial order $n$ and the columns correspond to the angular frequency $m$. Using  single-indexed Zernike polynomials, we rewrite Eq.~\eqref{eq:upphi} in polar coordinates as 
\begin{equation}
    \upphi(\rho ,\varphi )=\sum_q a_qZ_q(\rho ,\varphi ).
    \label{eq:upphi1}
\end{equation}

Due to the relationships between the intensity image $I(x,y)$ in Eq.~\eqref{eq:I(x,y)}, the PSF $h(x,y)$ in Eq.~\eqref{eq:h(x,y)}, and the Zernike coefficients $a_q$ in Eq.~\eqref{eq:upphi1}, the Zernike coefficients $a_q$ correspond to a particular intensity result and can thus be used to parameterize and model the associated PSF of an aberrated imaging system.

Unfortunately, there is an ambiguity associated with determining the Zernike coefficients from an intensity image alone without knowledge of the field phase. In particular, there are some Zernike polynomials $Z_q$ which generate the same PSF intensity image for oppositely-signed Zernike coefficients (details in Appendix). Due to this ambiguity, neural networks and other learning machines which have been extensively used to estimate PSFs for aberrated imaging systems \cite{jin2018machine, paine2018machine, nishizaki2019deep, zhang2019machine, tian2019dnn} face difficulty in directly predicting Zernike coefficients from intensity images \cite{lu2022mitigating, nishizaki2019deep}. In \cite{nishizaki2019deep}, the deep learning model completely fails to predict Zernike coefficients for the in-focus setup from an extended source object due to this intensity image ambiguity (although the ambiguity is not pointed out or identified as the source of failure by the authors). In an attempt to improve performance, the authors in \cite{nishizaki2019deep} analyze different preconditioners: overexposure, defocus, and scatter, and assess the impact on Zernike coefficient prediction. In \cite{lu2022mitigating}, the authors recognize the PSF intensity image ambiguity and address it by performing multiple measurements (focused and defocused intensity images) of the source object, passing them through a feature extractor block followed by a neural network (Resnet50).

In our work, we propose a novel approach to address this ambiguity. We begin by identifying the Zernike polynomials which are susceptible to the ambiguity and for these polynomials we only consider the prediction of the absolute value of the associated Zernike coefficients. For the remainder of this work, we will refer to the signless Zernike coefficients for polynomials susceptible to the ambiguity in addition to the signed Zernike coefficients not susceptible to the ambiguity collectively as the \textit{modified Zernike coefficients}. We use a deep neural network to predict the modified Zernike coefficients directly from intensity images. We show that these predicted coefficients can then be used to model the PSF of the aberrated imaging system. Finally, we consider the performance of our network in predicting the modified Zernike coefficients for both point source and extended source objects using incoherent imaging scenarios with both low and high noise levels (Poisson noise and read-out noise). Note that the proposed method is not a complete wavefront sensing approach, particularly because it does not consider the sign of the Zernike coefficients for polynomials that are susceptible to ambiguity. However, it works unambiguously for recovering the point spread function and provides a new construct and insight for the development of novel wavefront sensing methods \cite{zhu2022adaptive, guo2019improved, li2022prediction, shohani2023using, guo2022adaptive}.

The remainder of the work is organized as follows. In Section \ref{sec:Methods}, we describe the methodology, data generation process, and deep learning architecture to predict modified Zernike coefficients. In Section \ref{sec:Results}, we illustrate our results and discuss the logical interpretation of these results. In Section \ref{sec:Conclusion}, we conclude our work.

\section{Methods}
\label{sec:Methods}

\subsection{Methodology}

As has been pointed out in Section \ref{sec:RepresentingWavefrontDistortions}, there is an ambiguity associated with predicting the Zernike coefficients from intensity images. In particular, we have found via symmetry properties of the Fourier transform \cite{bracewell1986fourier} (details in Appendix) that angularly even Zernike polynomials which correspond to even values of radial order $n$ and even values of angular frequency $m$ generate the same PSF and thus the same intensity image for oppositely signed Zernike coefficients. In other words, the signs of these coefficients are immaterial for the description of the intensity PSF. Fig.~\ref{tbl:ZernikeTable} shows the first 28 polynomials where labels of angularly even polynomials are colored \textcolor{myFireBrick}{red}. To address the ambiguity, we predict the signless Zernike coefficients for angularly even polynomials and the signed Zernike coefficients for angularly odd polynomials using a deep neural network. This approach might not be sufficient for applications such as wavefront sensing where the coefficient signs are important for specifying the actual shape of the wavefront, but it is useful for characterizing and correcting intensity images of an aberrated imaging system.

\subsection{Data Generation}
\label{subsec:Data Generation}

A dataset was generated using the imaging model and Zernike polynomial representation of the phase aberration given in Section \ref{sec:Introduction}. More specifically, the wavefront phase distortion due to the aberrated system was introduced in the pupil plane by fitting a Kolmogorov turbulence phase screen that is applied to the pupil. The strength of the phase screen was parameterized by the ratio $D/r_0$, where $D$ is the diameter of the pupil and $r_0$ is the Fried parameter (the coherence diameter of the transverse turbulence field) \cite{underwood2013wave, zhan2019wave}. Other simulation values include the wavelength $\lambda = 0.5~\mu\mathrm{m}$ and pupil diameter $D=0.4~\mathrm{m}$. 

\begin{figure*}[tbh]
    \centering
     \includegraphics[width=\linewidth]{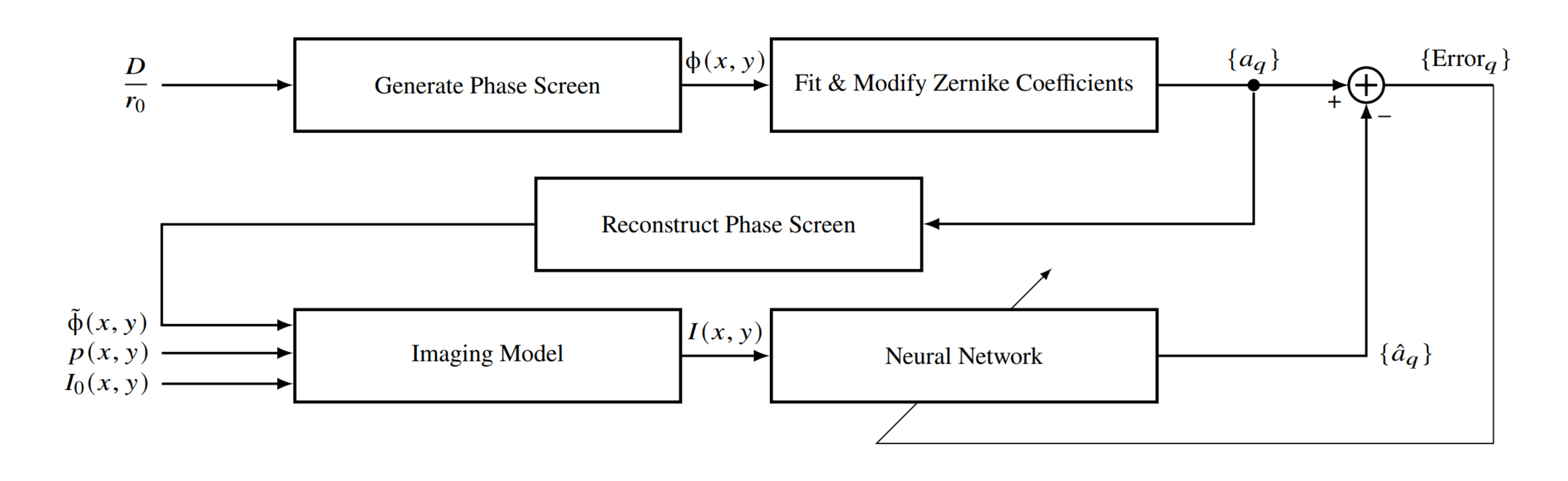}
    \caption{Block diagram of the proposed training methodology.}
    \label{fig:BlockDiagram}
\end{figure*}

\begin{figure*}[tbh!]
    \centering
  	\begin{minipage}[b]{0.5\linewidth}
  		\centering
  		\subfigure[]{
  		{\includegraphics[trim = 110 40 90 25, clip, width = 0.97\linewidth] {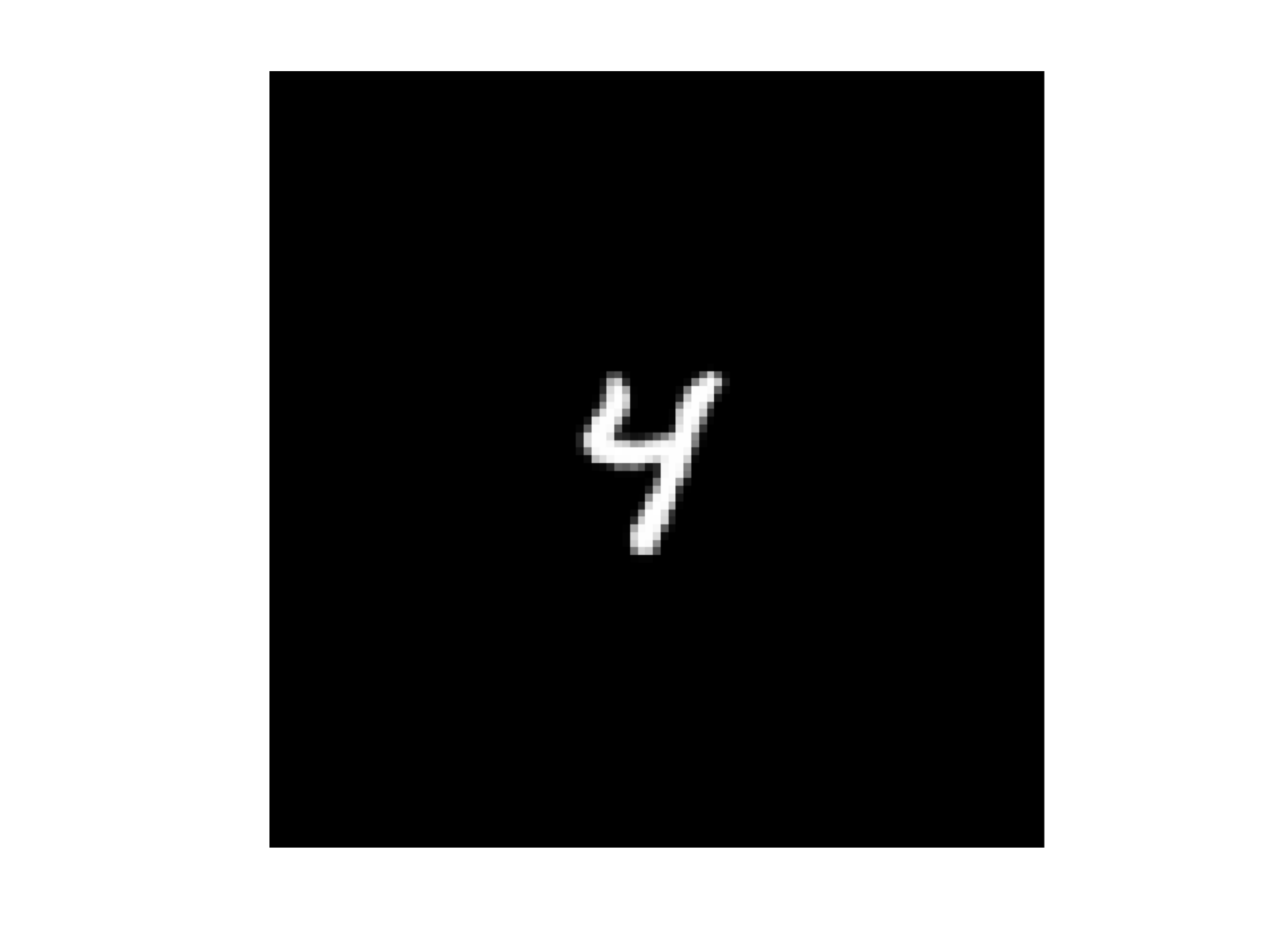}}
  		\label{fig:emnist_image}
  	}
  	\end{minipage}\begin{minipage}[b]{0.5\linewidth}
  		\centering
  		\subfigure[]{
  		{\includegraphics[trim = 110 40 90 25, clip, width = 0.97\linewidth]{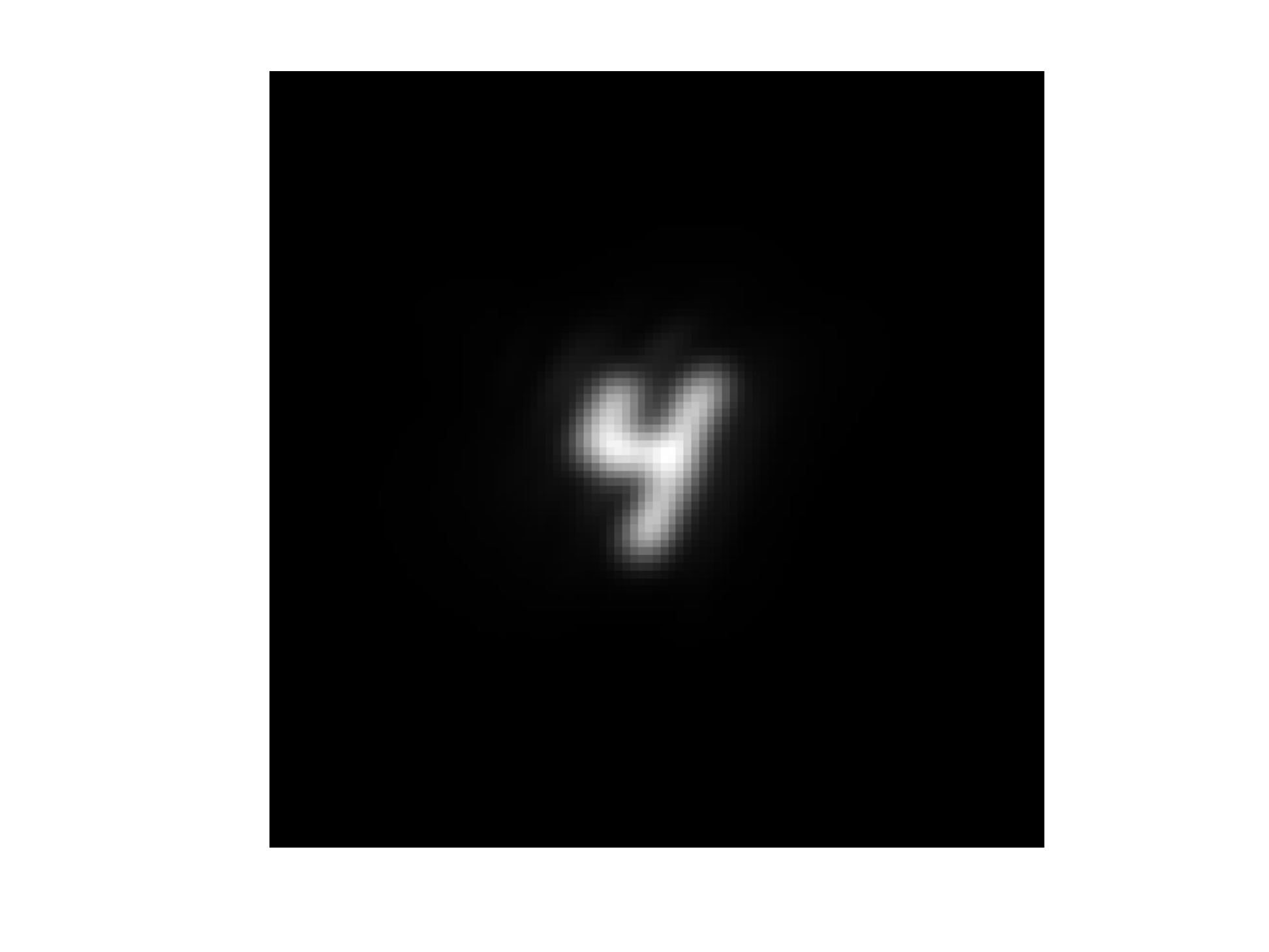}}
  		\label{fig:D/r0=2}
  	}
  	\end{minipage}
        \begin{minipage}[b]{0.5\linewidth}
  		\centering	
  		\subfigure[]{
  		{\includegraphics[trim = 110 40 90 25, clip, width = 0.97\linewidth]{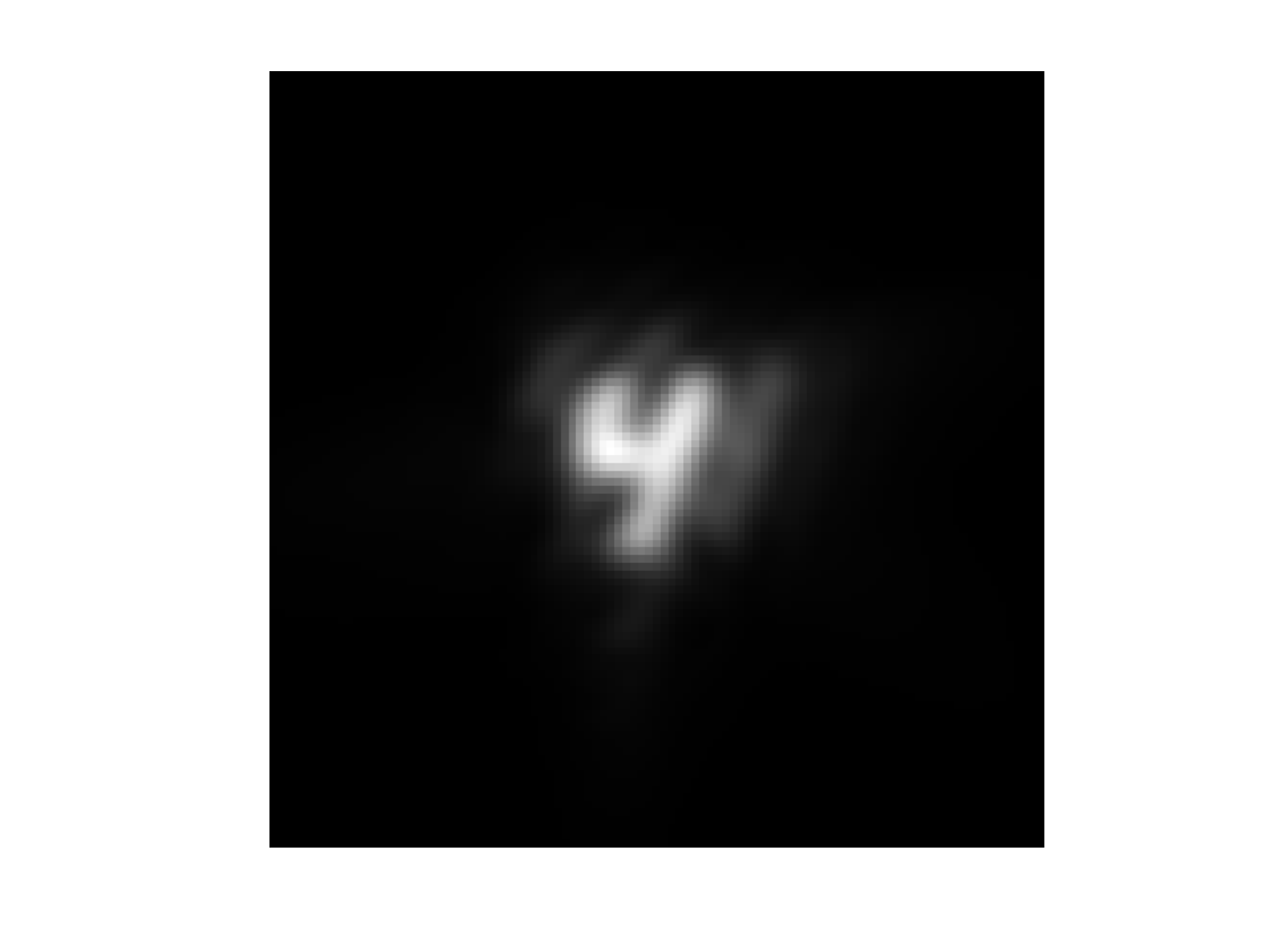}}
  		\label{fig::D/r0=5}
  	}
  	\end{minipage}\begin{minipage}[b]{0.5\linewidth}
  		\centering
  		\subfigure[]{
  		{\includegraphics[trim = 110 40 90 25, clip, width = 0.97\linewidth]{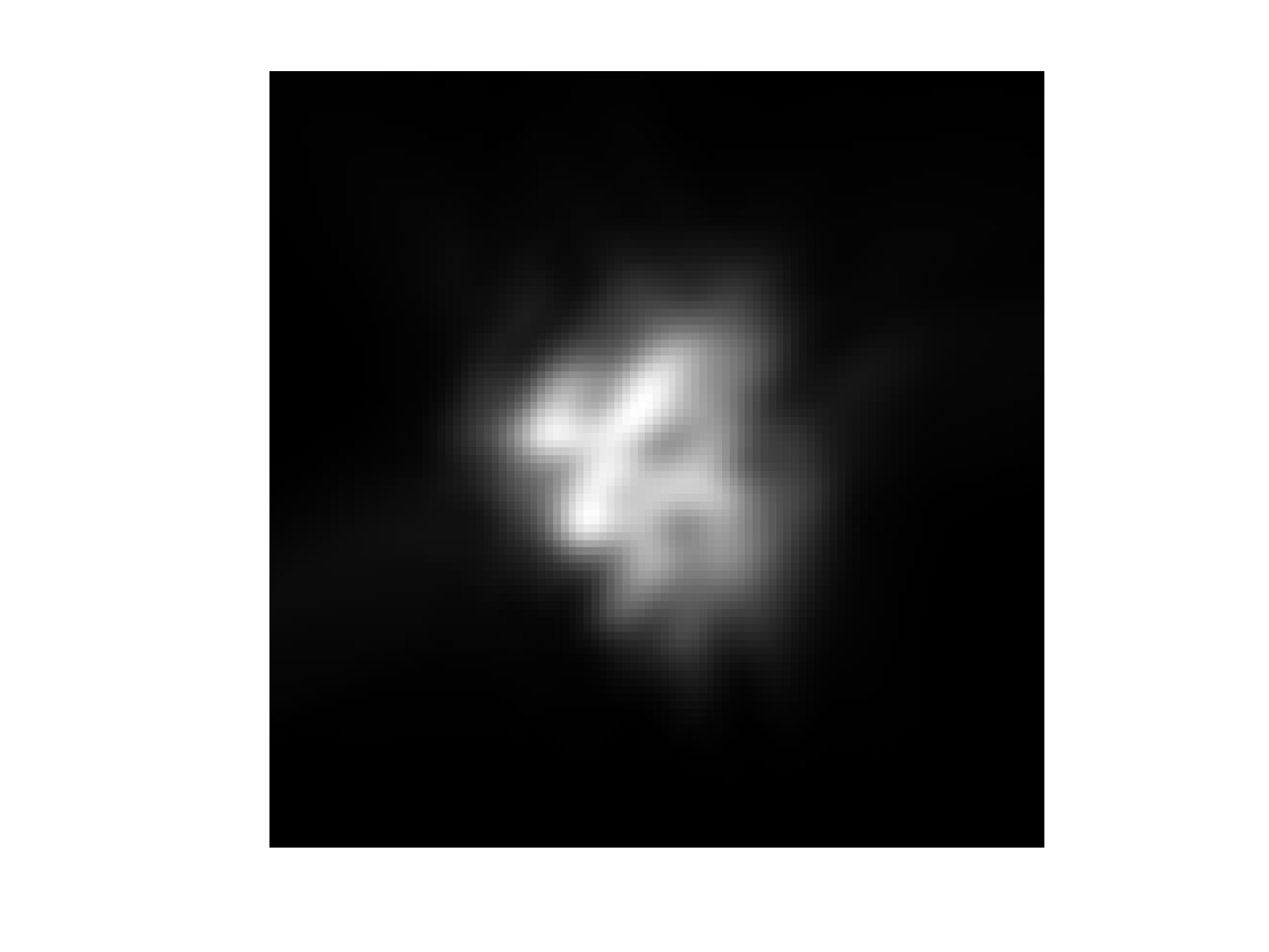}}
  		\label{fig::D/r0=10}
  	}
  	\end{minipage}
    \caption{Example intensity images of an extended source object for different $\frac{D}{r_0}$ ratios: (a) source object (b)-(d) intensity images for $\frac{D}{r_0}=2, 5, 10$ respectively. The frames shown are $100\times100$ pixels.}
    \label{fig:degraded images}
\end{figure*}

A block diagram of our training methodology is given in Fig.~\ref{fig:BlockDiagram}. The process begins with the generation of a random phase screen $\upphi(x,y)$ that obeys the Kolmogorov spectrum for a selected $D/r_0$ value \cite{schmidt2010numerical}. The first 28 Zernike coefficients of the phase screen $\upphi(x,y)$ are fitted using matrix inversion \cite{Christopher2023}. Other authors have demonstrated the effective estimation of this number of coefficients for turbulence-degraded wavefronts \cite{lu2022mitigating}. Then, the coefficients are modified by discarding the signs of the coefficients corresponding to angularly even Zernike polynomials resulting in the modified Zernike coefficients $\{a_q\}$. Additionally, the first three Zernike coefficients are considered as zero because there is no relationship between the $Z_0$ (piston) polynomial and the PSF intensity image \cite{jin2018machine}, and the $Z_1$ (tip) and $Z_2$ (tilt) terms correspond to simple offsets in the image plane that can be easily found using centroiding algorithms or other registration methods \cite{paine2018machine, delabie2014accurate}. The modified phase screen $\tilde{\upphi}(x,y)$ is reconstructed and used along with the pupil function $p(x,y)$ and source object $I_0(x,y)$ in the imaging model described by Eq.~\eqref{eq:I(x,y)} to generate the intensity image $I(x,y)$ for neural network input. Finally, the modified Zernike coefficients $\{a_q\}$ are compared to the output of the neural network $\{\hat{a}_q\}$ to determine the prediction error $\{\text{Error}_q\}$ associated with each Zernike coefficient. As an example, Fig.~\ref{fig:degraded images} shows a source object $I_0(x,y)$ and three intensity images $I(x,y)$ corresponding to three different $D/r_0$ values. It is noted that it is possible to skip the phase screen generation step and use random draws of Zernike coefficient values subject to known variances corresponding to the Kolmogorov spectrum \cite{noll1976zernike} but we find it useful to have a high spatial resolution version of the original phase screen for comparison purposes. 

Four different scenarios at nine atmospheric turbulence strengths $D/r_0 = 2, 3, \ldots, 10$ were considered. For each scenario, $54000$ intensity images ($512\times512$ pixels, $6000$ images per $D/r_0$ value) were generated and then partitioned $45000/4500/4500$ for training/validation/testing purposes. The first scenario considers only a point source object with zero noise. The remaining three scenarios consider extended source objects at zero/low/high noise levels, respectively. More specifically, $6000$ different extended source objects ($5000/500/500$ for training/validation/testing purposes) were used for each $D/r_0$. The extended source objects ($28\times28$ pixels) were taken from the EMNIST database \cite{cohen2017emnist} and were zero-padded to $512\times512$ pixels. These objects have relatively simple spatial features, which provide the opportunity to extract the PSF even in the presence of wavefront components due to the object. For the low noise scenario, Poisson noise with peak photon levels of $4000$ and read out noise with zero mean, standard deviation $10$ was used. For the high noise scenario, Poisson noise with peak photon levels of $15000$ and read out noise with zero mean, standard deviation $100$ was used. These noise levels are similar to those used in \cite{paine2018machine}.

\subsection{Model Architecture}
\begin{figure*}[tbh!]
    \centering
    \includegraphics[width=\linewidth]{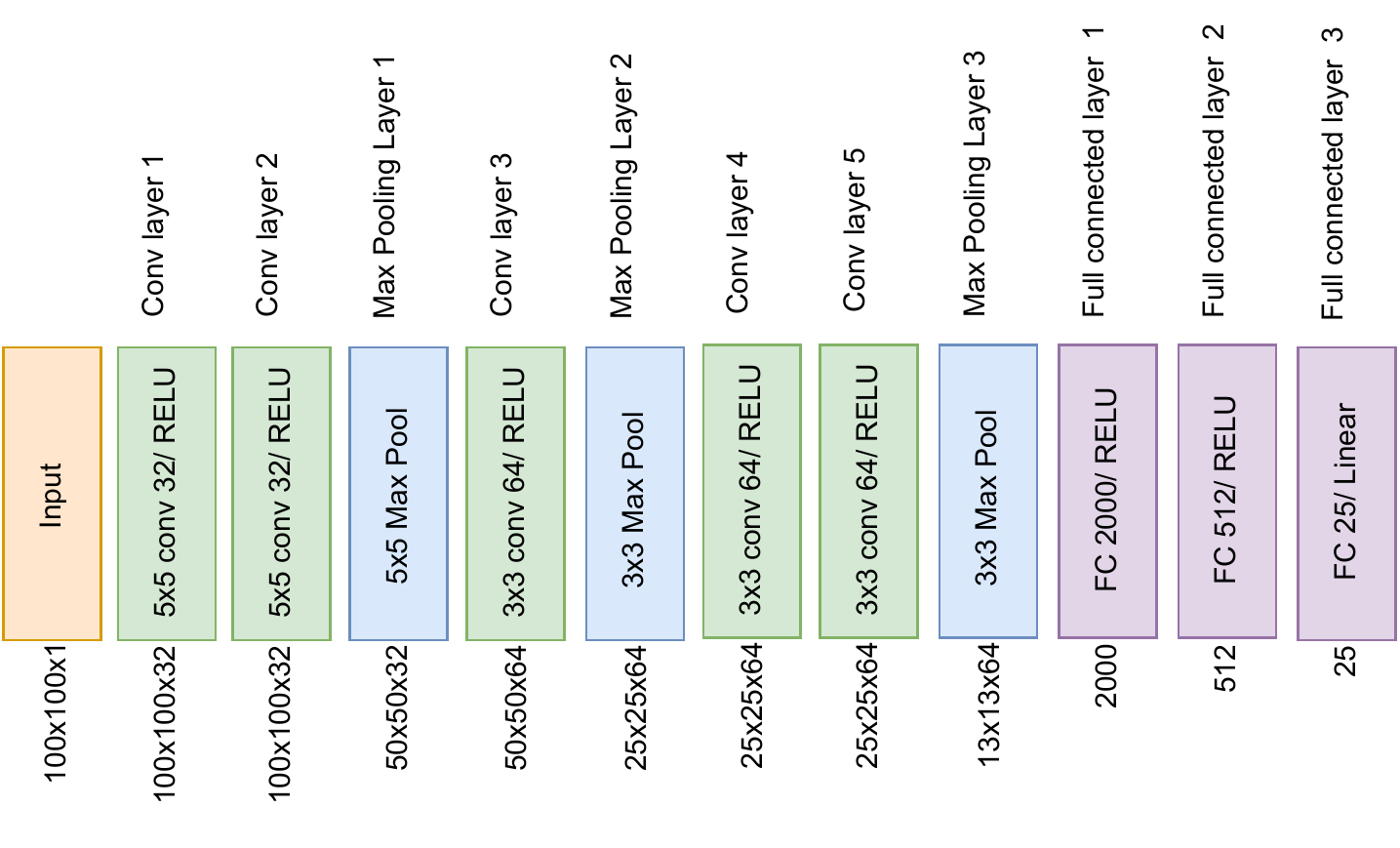}
    \caption{The neural network architecture used for the estimation consists of several convolutional and pooling layers followed by three fully connected layers. The architecture is based on AlexNet \cite{hu2020simplifying, krizhevsky2012imagenet}.}
    \label{fig:AlexNet}
\end{figure*}

We use a deep neural network architecture based on AlexNet \cite{hu2020simplifying, krizhevsky2012imagenet} to predict the modified Zernike coefficients from PSF intensity images. The model structure is illustrated in Fig.~\ref{fig:AlexNet}. The model is implemented in the Julia language \cite{bezanson2012julia} using the Flux package \cite{innes2018flux}. The $512\times512$ pixel PSF intensity images for the data set are centrally cropped down to $100\times100$ pixels and the amplitude is normalized using min-max normalization \cite{hu2020simplifying} prior to input to the neural network. The cropping procedure is carried out to exclude the outer zero-value pixels that do not contain any significant information for further analysis. The model takes cropped ($100\times100$ pixels) and amplitude normalized (in range $[0,1]$) PSF images as input and outputs predictions for the modified Zernike coefficients that model the PSF of the aberrated imaging system. The network consists of $5$ convolution layers, $3$ max-pooling layers, and $3$ fully-connected layers. Convolution layers $1$ and $2$ use $32$ kernels with kernel size $5\times5$ and convolution layers $3$, $4$ and $5$ use $64$ kernels with kernel size $3\times3$. Each convolution layer in the model is followed by a RELU activation. Max pooling layer $1$ performs a $5\times5$ kernel operation and max pooling layers $2$ and $3$ perform $3\times3$ kernel operations. The fully-connected layers $1$ and $2$ are followed by a RELU activation and have $2000$ and $512$ nodes, respectively. The final fully-connected layer has $25$ nodes followed by a linear activation. We use mean squared error (MSE)  as the loss function and the ADAM optimizer \cite{kingma2014adam} with default settings (learning rate $\upeta = 0.001$ and decay rates $\mathcal{B} = (0.9, 0.999)$) as an optimizer for the model. The neural network is trained for $20$ epochs with a batch size of $100$ for each scenario described earlier as in Section \ref{subsec:Data Generation}. During the training process, the validation MSE typically levels off but does not diverge and the neural network with the lowest MSE for the validation data is saved for each scenario.

\section{Results \& Discussion}
\label{sec:Results}
\begin{figure*}[t!]
    \centering
     \includegraphics[trim = 30 5 45 30, clip, width = 0.95\linewidth]{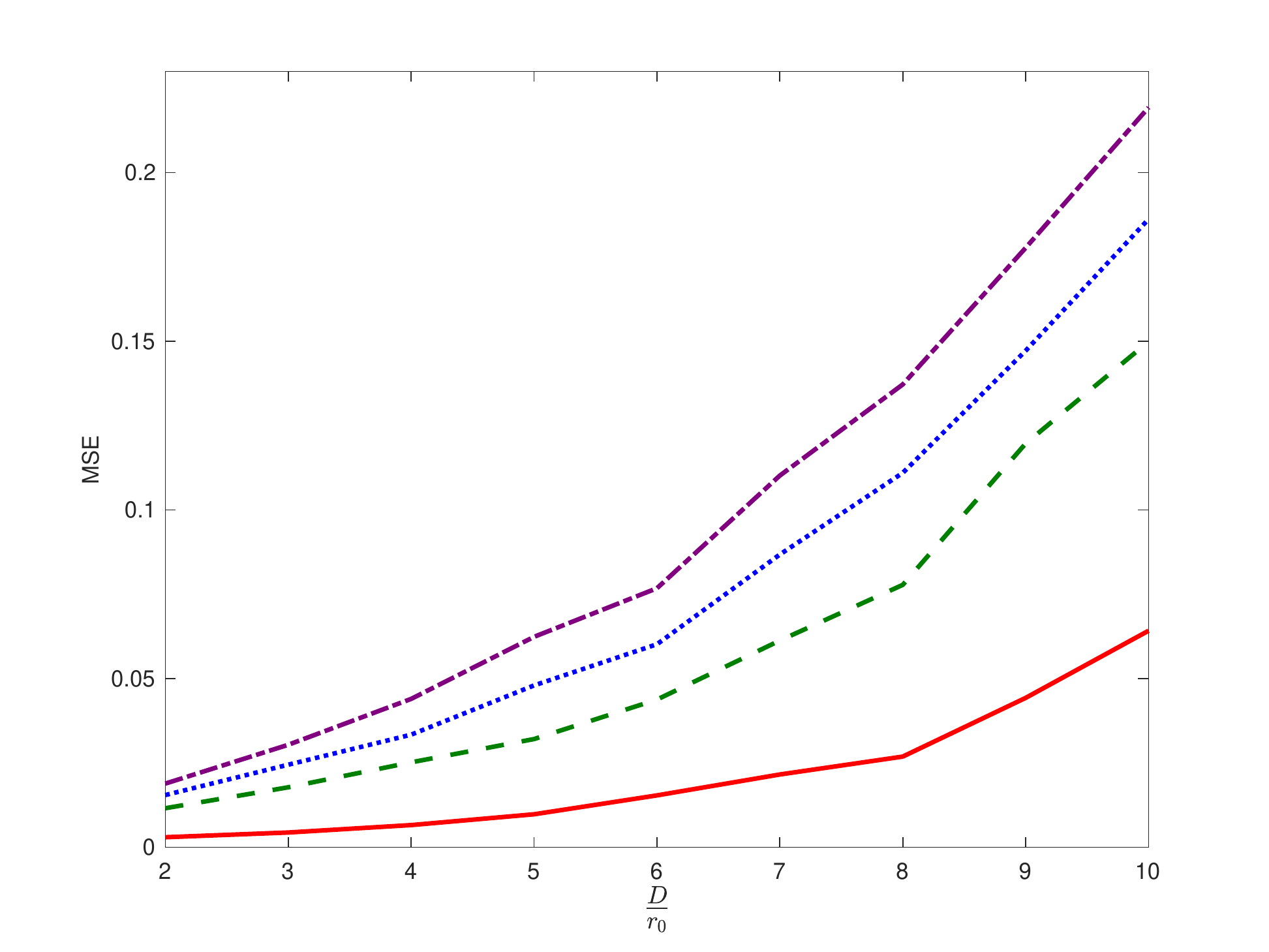}
    \caption{The MSE as a function of ${D}/{r_0}$ for ({\protect\textcolor{myRed}{\solidMrule[10mm]}}) a point source object with no noise, ({\protect\textcolor{myGreen}{\dashedruleM}}) extended source objects with no noise, ({\protect\textcolor{myBlue}{\dotruleM}}) extended source objects with low noise, and ({\protect\textcolor{myPurple}{\dashdotruleM}})  extended source objects with high noise}
    \label{fig:MSE}
\end{figure*}

\begin{figure*}[tbh!]
    \centering
  	\begin{minipage}[b]{0.5\linewidth}
  		\centering
  		\subfigure[]{
  		{\includegraphics[trim = 110 250 90 210, clip, width = 0.75\linewidth] {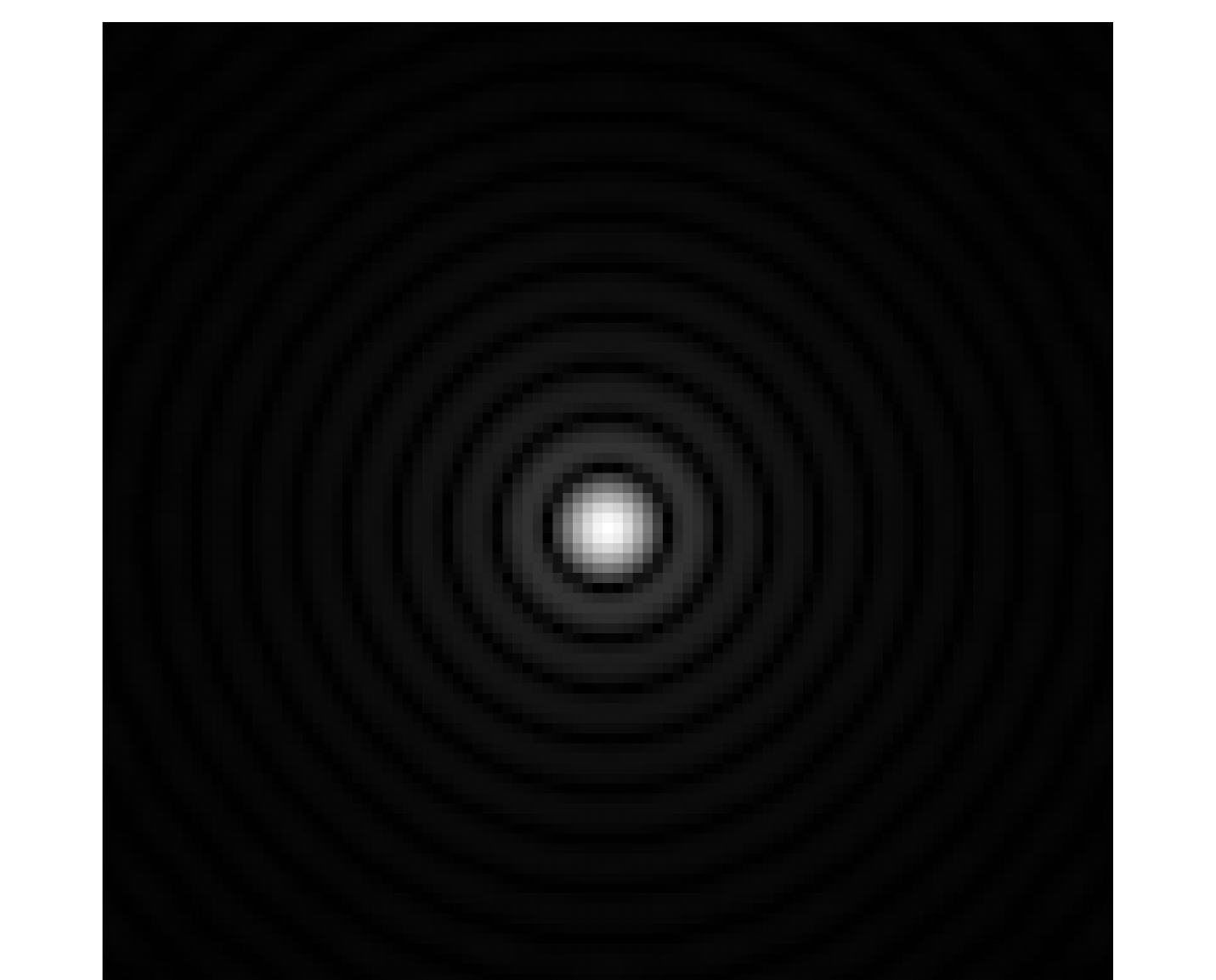}}
  		\label{fig:point_source1}
  	}
  	\end{minipage}
        \begin{minipage}[b]{0.5\linewidth}
  		\centering	
  		\subfigure[]{
  		{\includegraphics[trim = 110 210 90 190, width = 0.75\linewidth]{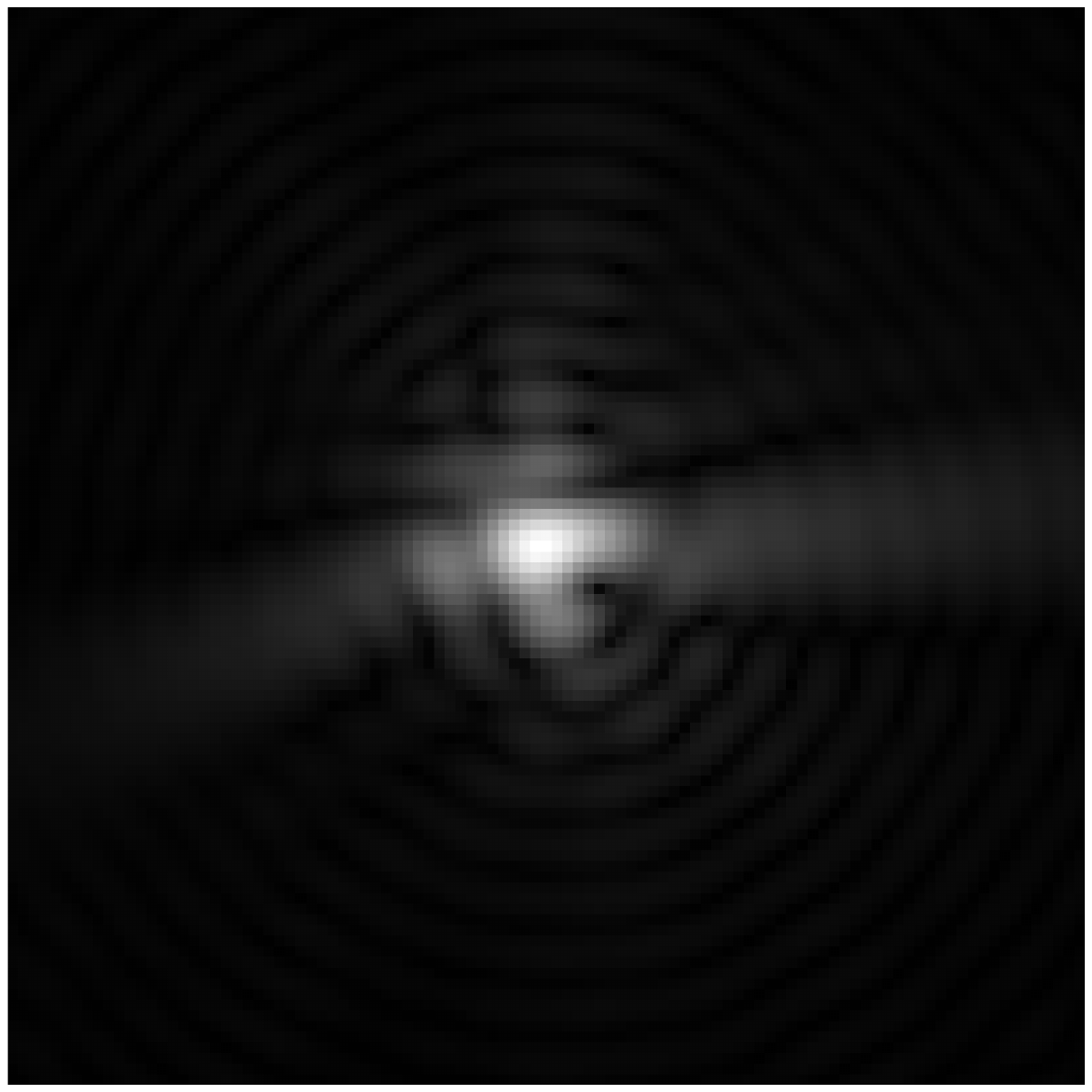}}
  		\label{fig:point_PSF1}
  	}
  	\end{minipage}\begin{minipage}[b]{0.5\linewidth}
  		\centering
  		\subfigure[]{
  		{\includegraphics[trim = 110 210 90 190, clip, width = 0.75\linewidth]{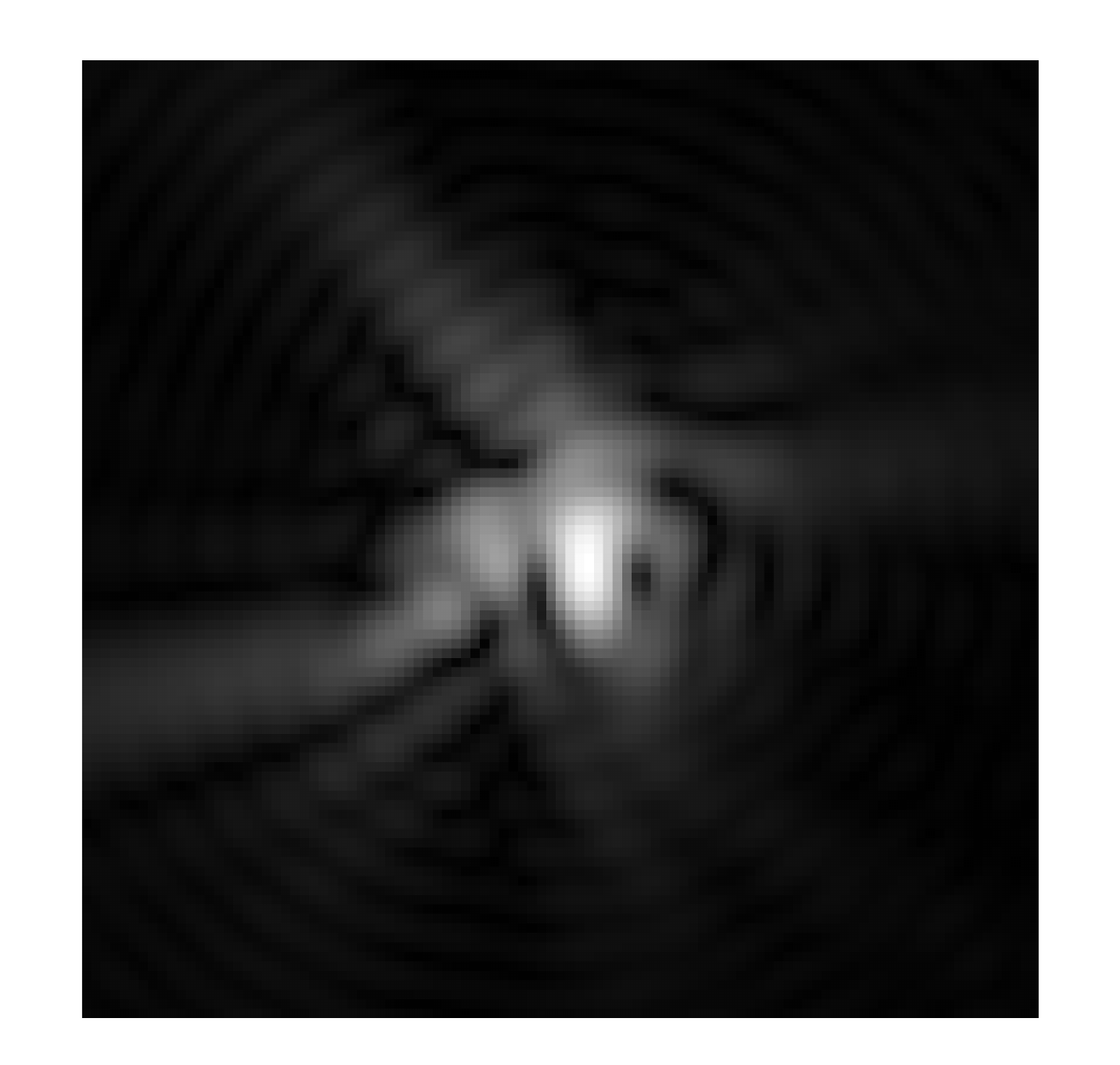}}
  		\label{fig:point_PSF2}
  	}
  	\end{minipage}
   \begin{minipage}[b]{1.0\linewidth}
  		\centering
  		\subfigure[]{
            {\includegraphics[width =0.8\linewidth]{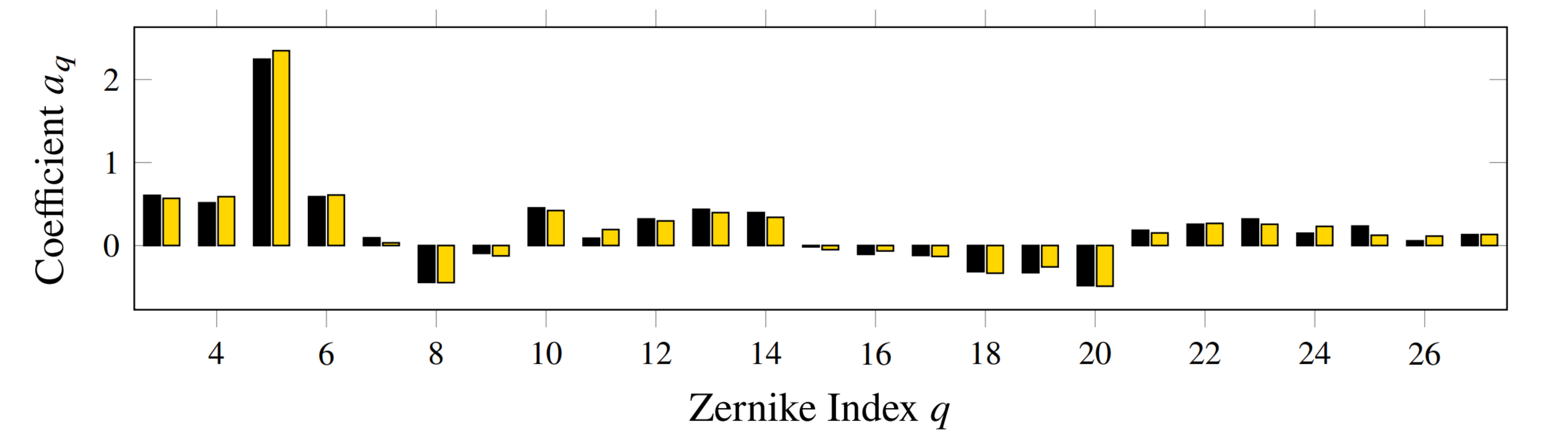}}
  		\label{fig:point_comp1}
  	}
  	\end{minipage}
   \begin{minipage}[b]{1.0\linewidth}
  		\centering
  		\subfigure[]{
            {\includegraphics[width =0.8\linewidth]{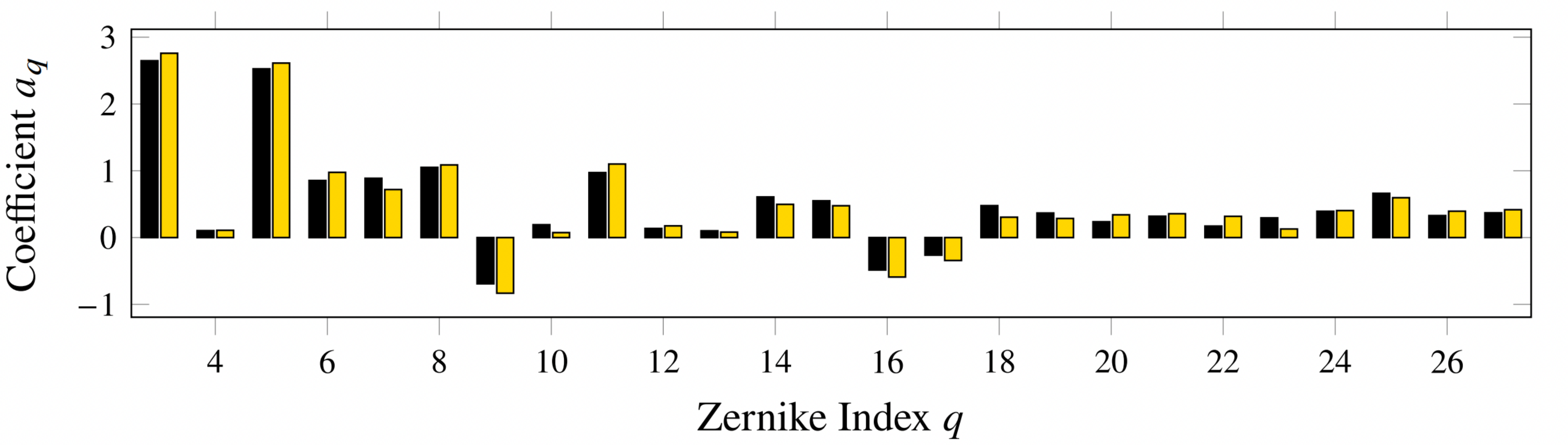}}
  		\label{fig:point_comp2}
  	}
  	\end{minipage}
   \caption{Demonstration of the proposed estimation for a point source object with zero noise. 
   (a) Diffraction limited PSF for the point source,
   (b) $I(x,y)$ for $D/r_0=2$,
   (c) $I(x,y)$ for $D/r_0=5$,
   (d) the actual (\textcolor{black}{\solidXXLrule[3.5mm]}) and predicted (\textcolor{myGold}{\solidXXLrule[3.5mm]}) Zernike coefficients corresponding to the intensity image in (b), and 
   (e) the actual (\textcolor{black}{\solidXXLrule[3.5mm]}) and predicted (\textcolor{myGold}{\solidXXLrule[3.5mm]}) Zernike coefficients corresponding to the intensity image in (c). The contrast was enhanced for display in (a)-(c).}
    \label{fig:Point_Coeff_Comparison}
\end{figure*}

\begin{figure*}[tbh!]
    \centering
  	\begin{minipage}[b]{0.5\linewidth}
  		\centering
  		\subfigure[]{
  		{\includegraphics[trim = 110 40 90 25, clip, width = 0.75\linewidth] {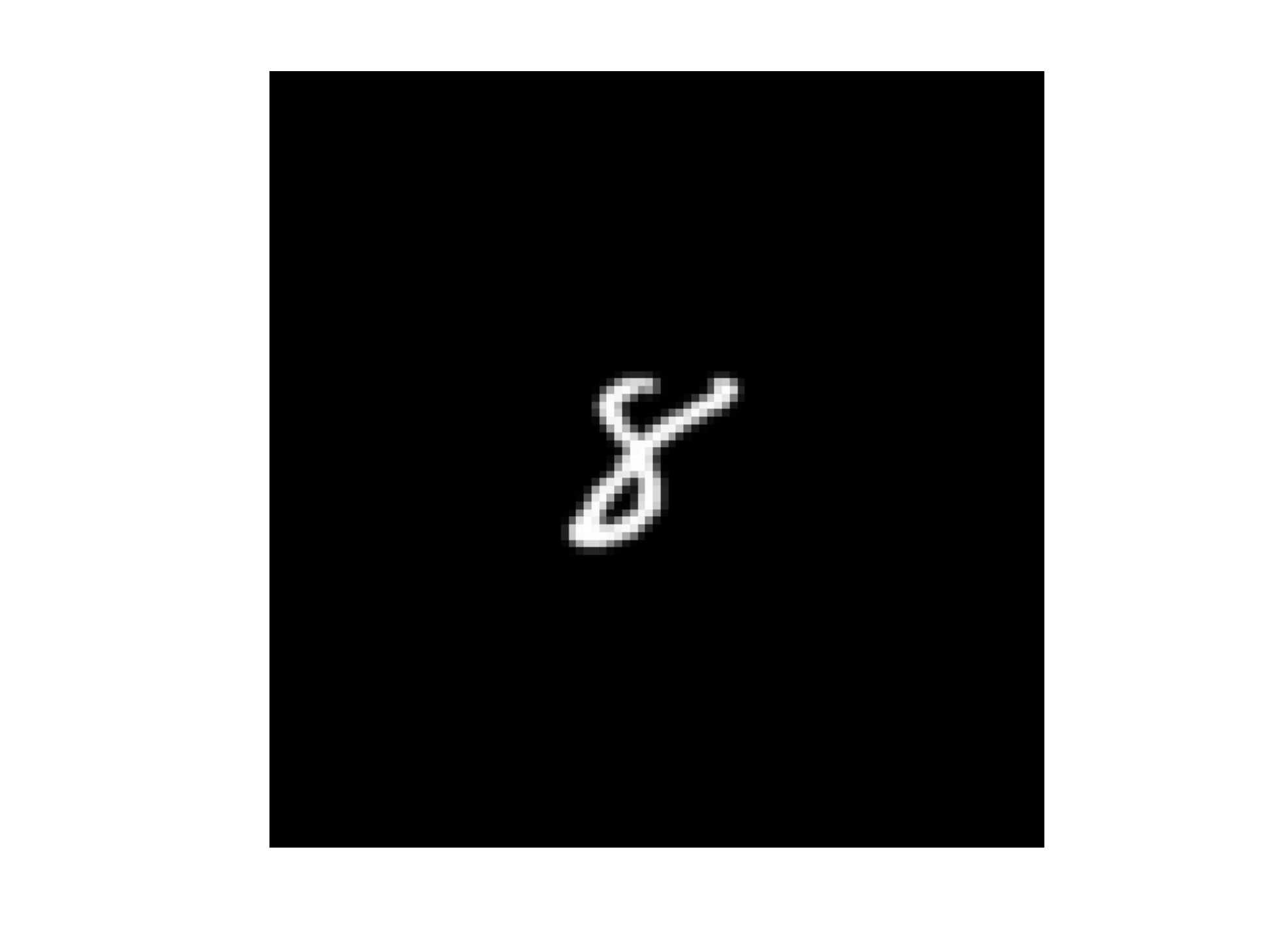}}
  		\label{fig:extended_source}
  	}
  	\end{minipage}
        \hfill\begin{minipage}[b]{0.5\linewidth}
  		\centering	
  		\subfigure[]{
  		{\includegraphics[trim = 110 40 90 25, clip, width = 0.75\linewidth]{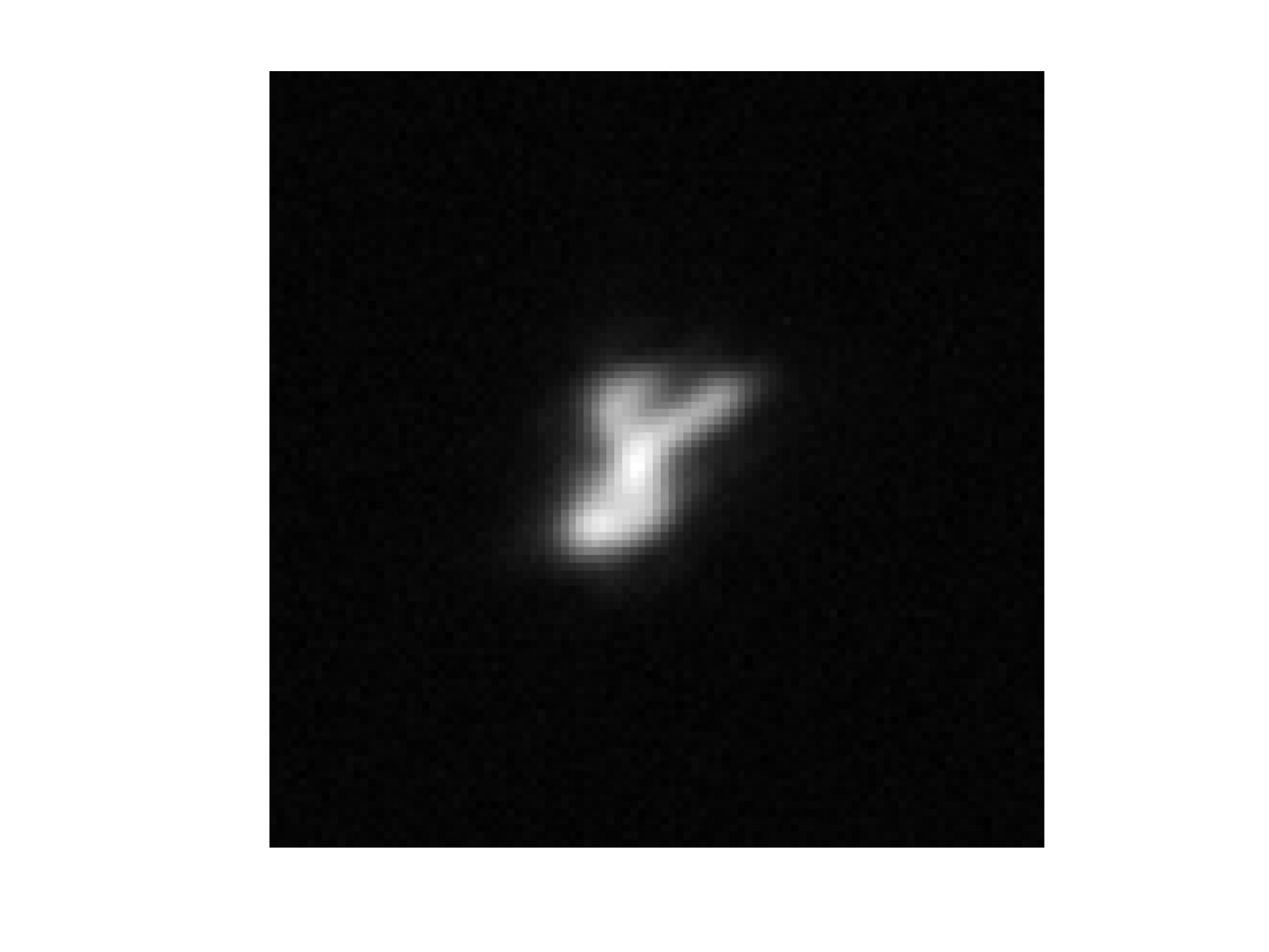}}
  		\label{fig:extended_PSF1}
  	}
  	\end{minipage}\hfill\begin{minipage}[b]{0.5\linewidth}
  		\centering
  		\subfigure[]{
  		{\includegraphics[trim = 110 40 90 25, clip, width = 0.75\linewidth]{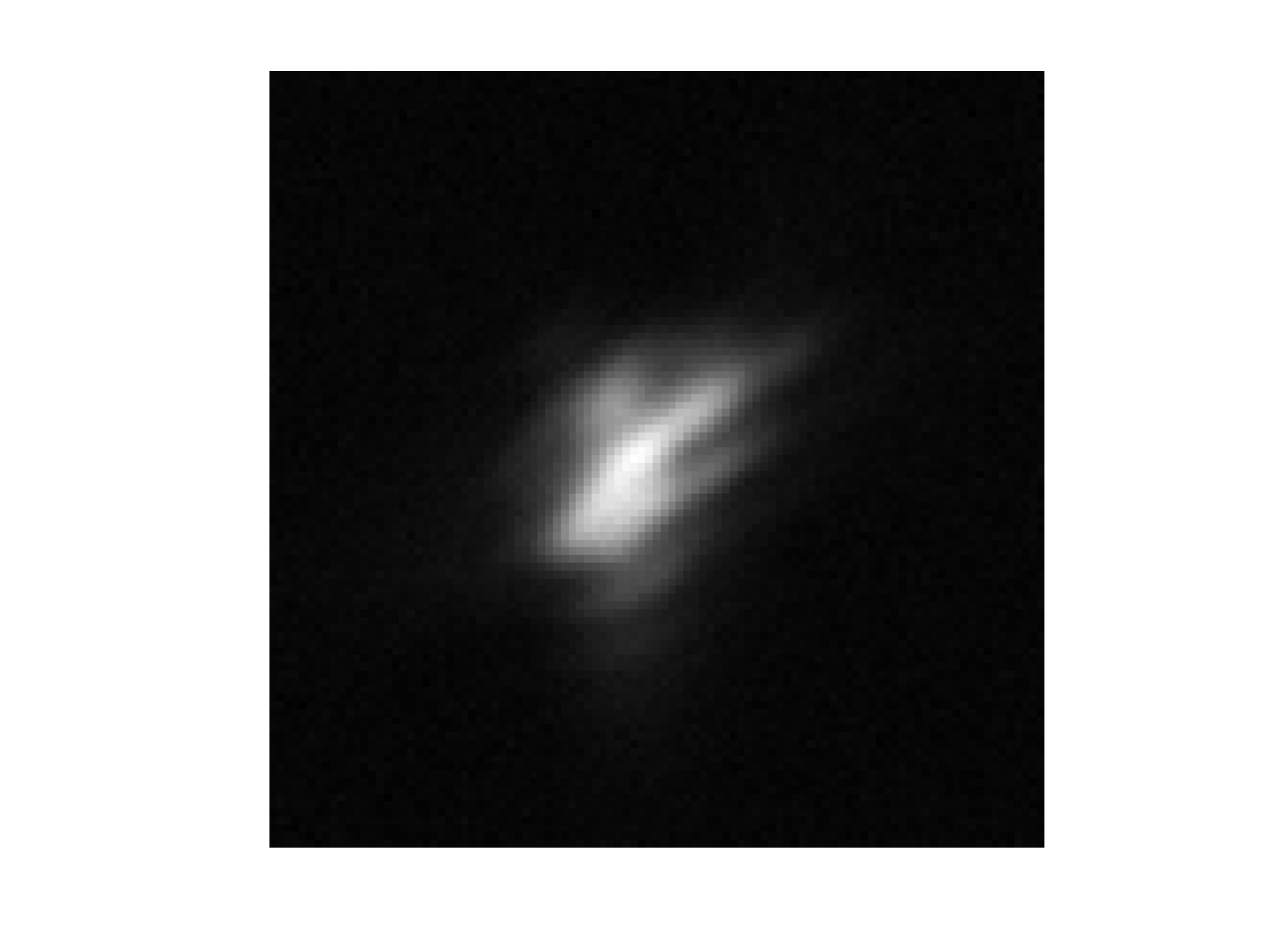}}
  		\label{fig:extended_PSF2}
  	}
  	\end{minipage}\hfill
   \begin{minipage}[b]{1.0\linewidth}
  		\centering
  		\subfigure[]{
            {\includegraphics[width =0.8\linewidth]{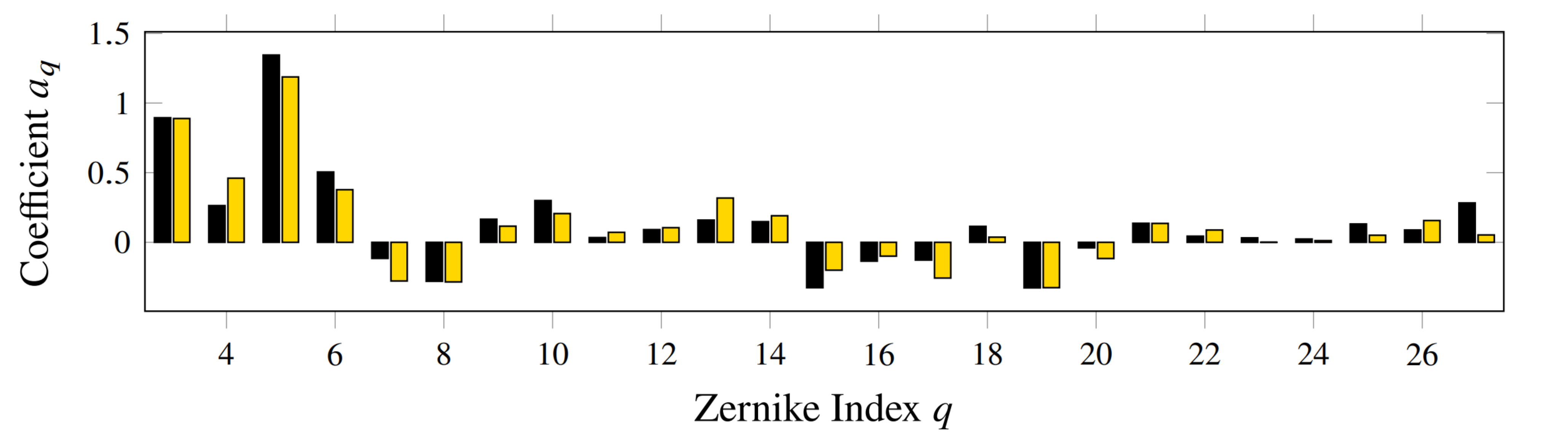}}
  		\label{fig:extended_comp1}
  	}
  	\end{minipage}
   \begin{minipage}[b]{1.0\linewidth}
  		\centering
  		\subfigure[]{
            {\includegraphics[width =0.8\linewidth]{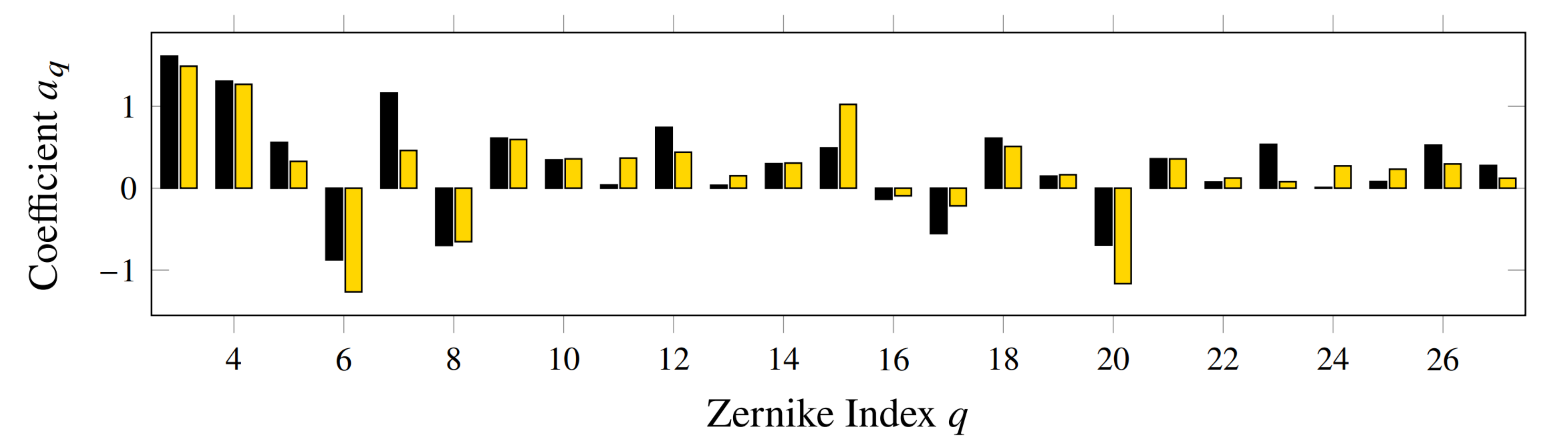}}
  		\label{fig:extended_comp2}
  	}
  	\end{minipage}
   \caption{Demonstration of the proposed estimation for an extended source object with high noise. 
   (a) $I_0(x,y)$ for the the extended source,
   (b) $I(x,y)$ for $D/r_0=2$,
   (c) $I(x,y)$ for $D/r_0=5$,
   (d) the actual (\textcolor{black}{\solidXXLrule[3.5mm]}) and predicted (\textcolor{myGold}{\solidXXLrule[3.5mm]}) Zernike coefficients corresponding to the intensity image in (b), and 
   (e) the actual (\textcolor{black}{\solidXXLrule[3.5mm]}) and predicted (\textcolor{myGold}{\solidXXLrule[3.5mm]}) Zernike coefficients corresponding to the intensity image in (c).}
    \label{fig:Extend_Coeff_Comparison}
\end{figure*}

Table \ref{tbl:summary} shows prediction results for each of the four scenarios described in Section \ref{subsec:Data Generation}. Each of these results provides an average MSE for predicting the modified Zernike coefficients from the testing PSF intensity images. Additionally, when we repeated scenario 1, but instead of the modified Zernike coefficients we tried to predict the signed Zernike coefficients for all polynomials, we only achieved an MSE of 0.5673. The results from Table \ref{tbl:summary} show that scenario 1 with a point source object and no noise produces the lowest average MSE whereas scenario 4 with the extended source objects set and high noise gives the largest MSE. This is expected because of the simplicity of the object and the lack of noise in scenario 1. The introduction of extended source objects without noise in scenario 2 causes an increase in the MSE compared to scenario 1. Scenario $3$ uses the same extended source objects as scenario $2$ but with low noise. The average MSE increases about $20\%$ compared to scenario 2. The presence of high noise in scenario $4$ with the extended source objects leads to a further increase of about $20\%$ in MSE. Fig.~\ref{fig:MSE} shows the average MSE for predicting modified Zernike coefficients from the testing PSF intensity images as a function of ${D}/{r_0}$ ratios for the four scenarios. The MSE grows exponentially as a function of ${D}/{r_0}$ for all scenarios, and the ratios between the MSE values of the different scenarios remain relatively constant. Fig.~\ref{fig:Point_Coeff_Comparison} illustrates the estimation results for a point source object with zero noise. The diffraction limited point source object is shown in Fig.~\ref{fig:point_source1} and the PSF intensity images  $I(x,y)$ for that point source with ${D}/{r_0}=2, 5$ are shown in Fig.~\ref{fig:point_PSF1} and Fig.~\ref{fig:point_PSF2} respectively. Fig.~\ref{fig:point_comp1} and Fig.~\ref{fig:point_comp2} display the actual and predicted Zernike coefficients corresponding to the intensity images in Fig.~\ref{fig:point_PSF1} and Fig.~\ref{fig:point_PSF2}, respectively. As can be seen in Fig.~\ref{fig:point_comp1} and Fig.~\ref{fig:point_comp2}, the difference between the actual and predicted Zernike coefficients is small for both ${D}/{r_0}$ cases for point source object at zero noise. Similarly, Fig.~\ref{fig:Extend_Coeff_Comparison} illustrates the estimation results for an extended source object with high noise. From Fig.~\ref{fig:extended_comp1} and Fig.~\ref{fig:extended_comp2}, it may be observed that the deviation between actual and predicted Zernike coefficients increases significantly for both ${D}/{r_0}$ cases for the extended source with high noise, compared to Fig.~\ref{fig:point_comp1} and Fig.~\ref{fig:point_comp2}. 

\begin{table}[]
\centering
\caption{Average MSE for the four scenarios described in Section \ref{subsec:Data Generation}.} 
\begin{tabular}{ccc}
\hline
\textbf{Scenario} & \textbf{Average MSE} & \textbf{Description}                            \\ \hline
1      & 0.0218      & Point Source Object                    \\ 
2      & 0.0599      & Extended Source Objects                 \\ 
3      & 0.0792      & Extended Source Objects with Low Noise  \\ 
4      & 0.0974      & Extended Source Objects with High Noise \\ \hline
\end{tabular}
\label{tbl:summary}
\end{table}

\section{Conclusion}
\label{sec:Conclusion}
A deep learning model based on a convolutional neural network was trained to predict modified Zernike coefficients in the pupil of an imaging system from a single turbulence-degraded intensity image. The modified Zernike coefficient set differs from a conventional set in that an absolute value is assigned to coefficients of even radial orders due to a sign ambiguity associated with using the intensity image. The modified set was shown to be sufficient for specifying the intensity PSF. Data for the learning model was created with an image simulation of a point object and simple extended objects for a range of turbulence and detection noise levels. The prediction MSE for the learning model shows that it is possible to recover a useful set of modified Zernike coefficients from an extended object intensity image subject to noise and turbulence. As expected, the results show that the point source object with no noise produces the lowest average MSE whereas the extended source objects with high noise give the largest MSE. In all cases, the MSE increases in a predictable way with turbulence strength ($D/r_0$). Future work could explore the prediction of higher order Zernike terms and the use of more varied source objects. The quality and utility of the PSFs derived from the predicted Zernike coefficients were not investigated in this work and are essential topics for future efforts.

\bigskip

\noindent\textbf{Funding} ~~~Office of Naval Research (N00014-21-1-2430).

\bibliographystyle{plainurl}
\bibliography{main}

\clearpage
\appendix

\section{Ambiguity Associated with Predicting Zernike coefficients from Intensity Images}
\label{sec:Ambiguity}

In order to explicitly show the ambiguity associated with predicting Zernike coefficients from intensity images, we utilize the symmetry properties of the Fourier transform \cite{bracewell1986fourier}. We begin by writing an expression for the point spread function $h(x,y)$ in terms of the Fourier transform as
\begin{equation}
    h(x,y)\propto \mathcal{F}\left\{p(x,y)\mathrm{e}^{\,\mathrm{j}\upphi(x,y)}\right\}\mathcal{F}^*\left\{p(x,y)\mathrm{e}^{\,\mathrm{j}\upphi(x,y)}\right\},
    \label{eq:h1(x,y)}
\end{equation}
and expressing the pupil function $p(x,y)$ and wavefront phase distortion $\upphi(x,y)$ in terms of real/imaginary and even/odd parts as
\begin{equation}
    p(x,y)=p_\textrm{re}(x,y)+ p_\textrm{ro}(x,y) + \mathrm{j}p_\textrm{ie}(x,y) + \mathrm{j}p_\textrm{io}(x,y)
    \label{eq:p1(x,y)}
\end{equation}
and
\begin{equation}
    \upphi(x,y)=\upphi_\textrm{re}(x,y)+ \upphi_\textrm{ro}(x,y) + \mathrm{j}\upphi_\textrm{ie}(x,y) + \mathrm{j}\upphi_\textrm{io}(x,y)
    \label{eq:upphi1(x,y)}
\end{equation}
where the subscripts $\textrm{re}$, $\textrm{ro}$, $\textrm{ie}$, and $\textrm{io}$ represent the real-even, real-odd, imaginary-even, and imaginary-odd parts, respectively. Under the assumption that the pupil is a real-even function then only term $p_\textrm{re}(x,y)$ in \eqref{eq:p1(x,y)} is non-zero. Furthermore, assuming the wavefront aberration phase is purely real, then only terms $\upphi_\textrm{re}(x,y)$ and $\upphi_\textrm{ro}(x,y)$ in \eqref{eq:upphi1(x,y)} are non-zero. Next, we consider the two special cases of particular interest. Specifically, the cases when the Zernike polynomial is either angularly even which corresponds to even radial order, or angularly odd which corresponds to odd radial order.
\subsection{Angularly Even Zernike Polynomial}
 An angularly even Zernike polynomial $Z_\mathrm{ae}(x,y)$ must be a real even symmetric function \cite{lakshminarayanan2011zernike}. Thus, for an angularly even Zernike polynomial $Z_\mathrm{ae}(x,y)$ the wavefront phase distortion $\upphi_\mathrm{re}(x,y)$ has real even symmetry. Therefore, for an angularly even polynomial $Z_\mathrm{ae}(x,y)$, we can write 
\begin{eqnarray}
    \mathcal{F}\left\{p(x,y)\mathrm{e}^{\,\mathrm{j}\upphi(x,y)}\right\}  &=& \mathcal{F}\left\{p_\mathrm{re}(x,y)\mathrm{e}^{\,\mathrm{j}\upphi_\mathrm{re}(x,y)}\right\}\\
     &=& \mathcal{F}\big\{p_\mathrm{re}(x,y)\cos{\upphi_\mathrm{re}(x,y)} + \mathrm{j} p_\mathrm{re}(x,y)\sin{\upphi_\mathrm{re}(x,y)}\big\}
    \label{eq:pos_coeff_ambiguity}
\end{eqnarray}
For Zernike coefficient $\pm a$, we have $\upphi_\mathrm{re}(x,y)=\pm aZ_\mathrm{ae}(x,y)$ and
\begin{eqnarray}
    \mathcal{F}\left\{p(x,y)\mathrm{e}^{\,\pm\mathrm{j} aZ_\mathrm{ae}(x,y)}\right\}  &=& \mathcal{F}\big\{p_\mathrm{re}(x,y)\cos[{\pm aZ_\mathrm{ae}(x,y)}] + \mathrm{j}p_\mathrm{re}(x,y)\sin[{\pm aZ_\mathrm{ae}(x,y)}]\big\}~~~~~~~~\\
    &=& \mathcal{F}\big\{p_\mathrm{re}(x,y)\cos[{aZ_\mathrm{ae}(x,y)}] \pm \mathrm{j}p_\mathrm{re}(x,y)\sin[{ aZ_\mathrm{ae}(x,y)}]\big\}\\
     &=& A(x,y) \pm \mathrm{j}B(x,y)
\end{eqnarray}
where $p_\mathrm{re}(x,y)\cos[{aZ_\mathrm{ae}(x,y)}]$ is a real even symmetric function (cosine of a real even symmetric function is also real even symmetric) and $\mathrm{j}p_\mathrm{re}(x,y)\sin[{ aZ_\mathrm{ae}(x,y)}]$ is an imaginary even symmetric function (sine of a real even symmetric function is also real even symmetric). Using symmetry properties of the Fourier transform, $A(x,y)=\mathcal{F}\big\{p_\mathrm{re}(x,y)\cos[{aZ_\mathrm{ae}(x,y)}] \big\}$ is real even and $\mathrm{j}B(x,y)=\mathcal{F}\big\{\mathrm{j}p_\mathrm{re}(x,y)\sin[{ aZ_\mathrm{ae}(x,y)}]\big\}$ is imaginary even. This leads to 
\begin{equation}
     h(x,y)\propto \left| \mathcal{F}\left\{p_\mathrm{re}(x,y)\mathrm{e}^{\,\mathrm{j}\upphi_\mathrm{re}(x,y)}\right\} \right|^2 = A^2(x,y) + B^2(x,y).
    \label{eq:h(x)_pos_ambiguity}
\end{equation}
Therefore, as can be seen from \eqref{eq:h(x)_pos_ambiguity},  an angularly even polynomial results in same value for $h(x,y)$ irrespective of the sign of the Zernike coefficient $a$, thus leading to an intensity image ambiguity.
\subsection{Angularly Odd Zernike Polynomial}
Similarly, an angularly odd Zernike polynomial $Z_\mathrm{ao}(x,y)$ must be a real odd symmetric function \cite{lakshminarayanan2011zernike}. Thus, for an angularly odd Zernike polynomial $Z_\mathrm{ao}(x,y)$ the wavefront phase distortion $\upphi_\mathrm{ro}(x,y)$ has real odd symmetry. Therefore, for an angularly odd polynomial $Z_\mathrm{ao}(x,y)$, we can write
\begin{eqnarray}
    \mathcal{F}\left\{p(x,y)\mathrm{e}^{\,\mathrm{j}\upphi(x,y)}\right\}  &=& \mathcal{F}\left\{p_\mathrm{re}(x,y)\mathrm{e}^{\,\mathrm{j}\upphi_\mathrm{ro}(x,y)}\right\}\\
    &=& \mathcal{F}\big\{p_\mathrm{re}(x,y)\cos{\upphi_\mathrm{ro}(x,y)} + \mathrm{j} p_\mathrm{re}(x,y)\sin{\upphi_\mathrm{ro}(x,y)}\big\}
    \label{eq:no_ambiguity}
\end{eqnarray}
For Zernike coefficient $\pm a$, we have $\upphi_\mathrm{ro}(x,y)=\pm aZ_\mathrm{ao}(x,y)$ and
\begin{eqnarray}
    \mathcal{F}\left\{p(x,y)\mathrm{e}^{\,\pm\mathrm{j} aZ_\mathrm{ao}(x,y)}\right\}  &=& \mathcal{F}\big\{p_\mathrm{re}(x,y)\cos[{\pm aZ_\mathrm{ao}(x,y)}] + \mathrm{j}p_\mathrm{re}(x,y)\sin[{\pm aZ_\mathrm{ao}(x,y)}]\big\}~~~~~~~~\\
    &=& \mathcal{F}\big\{p_\mathrm{re}(x,y)\cos[{ aZ_\mathrm{ao}(x,y)}] \pm \mathrm{j}p_\mathrm{re}(x,y)\sin[{ aZ_\mathrm{ao}(x,y)}]\big\}\\
     &=& C(x,y) \pm D(x,y)
\end{eqnarray}
where $p_\mathrm{re}(x,y)\cos[{aZ_\mathrm{ao}(x,y)}]$ is a real even symmetric function (cosine of a real odd symmetric function is real even symmetric) and $\mathrm{j}p_\mathrm{re}(x,y)\sin[{ aZ_\mathrm{ao}(x,y)}]$ is an imaginary odd symmetric function (sine of a real odd symmetric function is also real odd symmetric). Using symmetry properties of the Fourier transform, 
$C(x,y)=\mathcal{F}\big\{p_\mathrm{re}(x,y)\cos[{aZ_\mathrm{ao}(x,y)}] \big\}$
is real even and $D(x,y)=\mathcal{F}\big\{\mathrm{j}p_\mathrm{re}(x,y)\sin[{ aZ_\mathrm{ao}(x,y)}]\big\}$ is real odd. This leads to 
\begin{equation}
   h(x,y)\propto \left| \mathcal{F}\left\{p_\mathrm{re}(x,y)\mathrm{e}^{j\upphi_\mathrm{ro}(x,y)}\right\} \right|^2= \big[C(x,y) \pm D(x,y)\big]^2.
  \label{eq:no_ambiguity2}
\end{equation}
Therefore, as can be seen from \eqref{eq:no_ambiguity2}, an angularly odd polynomial results in a different value for $h(x,y)$ respective to the sign of the Zernike coefficient $a$, thus does not lead to an intensity image ambiguity.

\subsection{Example}
\begin{figure*}[t!]
    \centering
  	\begin{minipage}[b]{0.5\linewidth}
  		\centering
  		\subfigure[]{
  		{\includegraphics[trim = 110 40 90 25, clip, width = 0.97\linewidth] {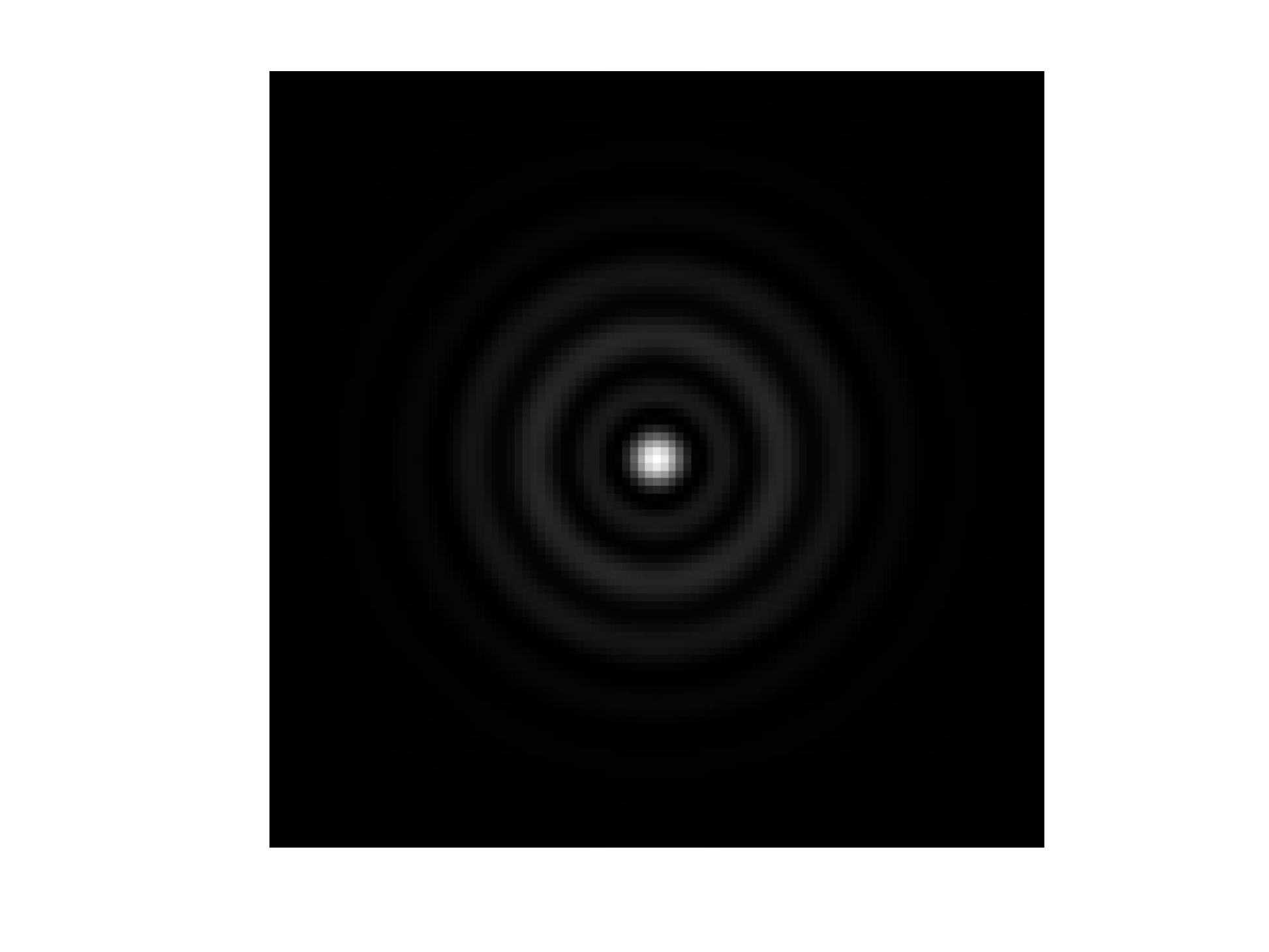}}
  		\label{fig:sig1A}
  	}
  	\end{minipage}\begin{minipage}[b]{0.5\linewidth}
  		\centering
  		\subfigure[]{
  		{\includegraphics[trim = 110 40 90 25, clip, width = 0.97\linewidth]{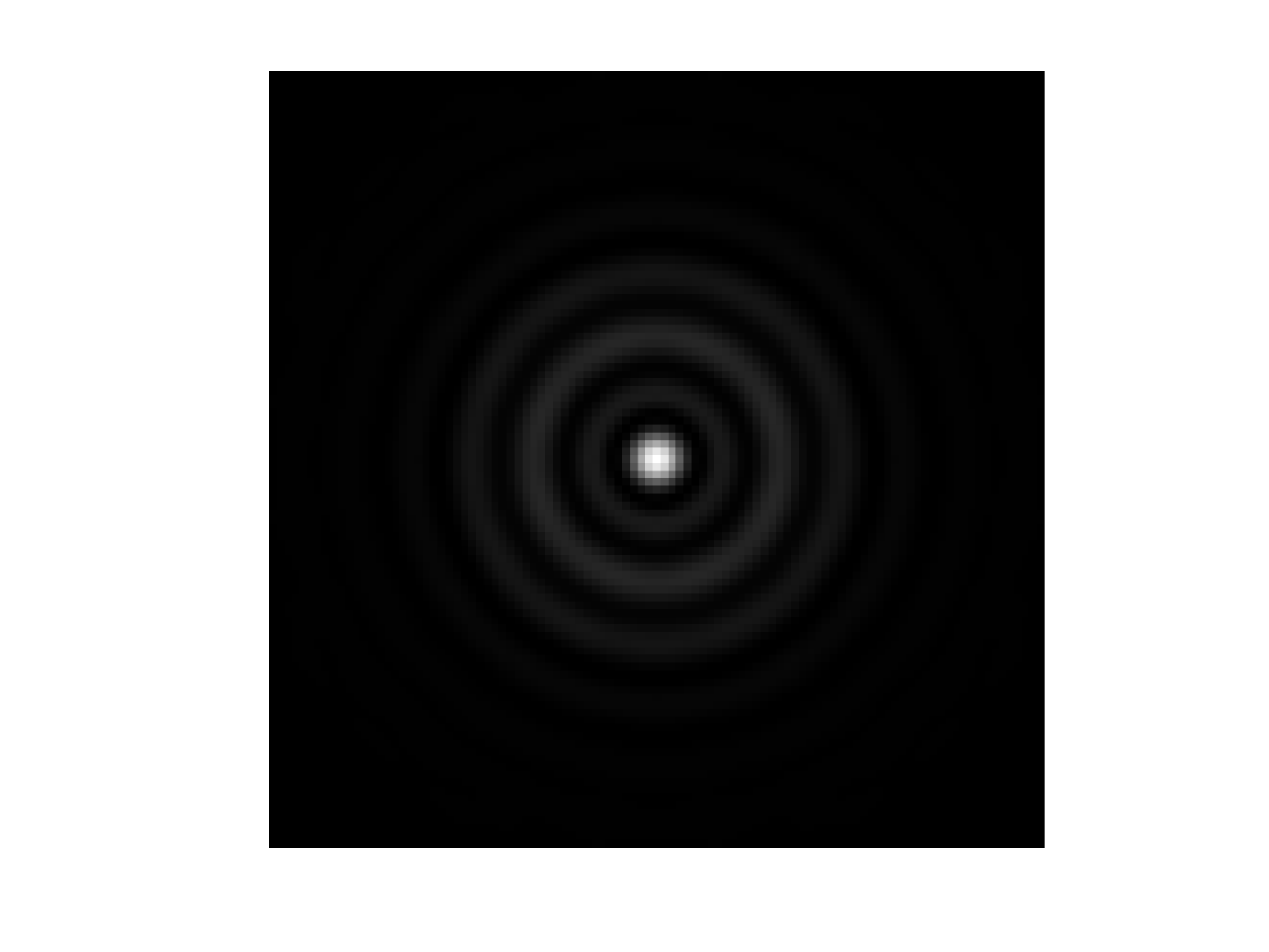}}
  		\label{fig:sig1B}
  	}
  	\end{minipage}
  	\begin{minipage}[b]{0.5\linewidth}
  		\centering	
  		\subfigure[]{
  		{\includegraphics[trim = 110 40 90 25, clip, width = 0.97\linewidth]{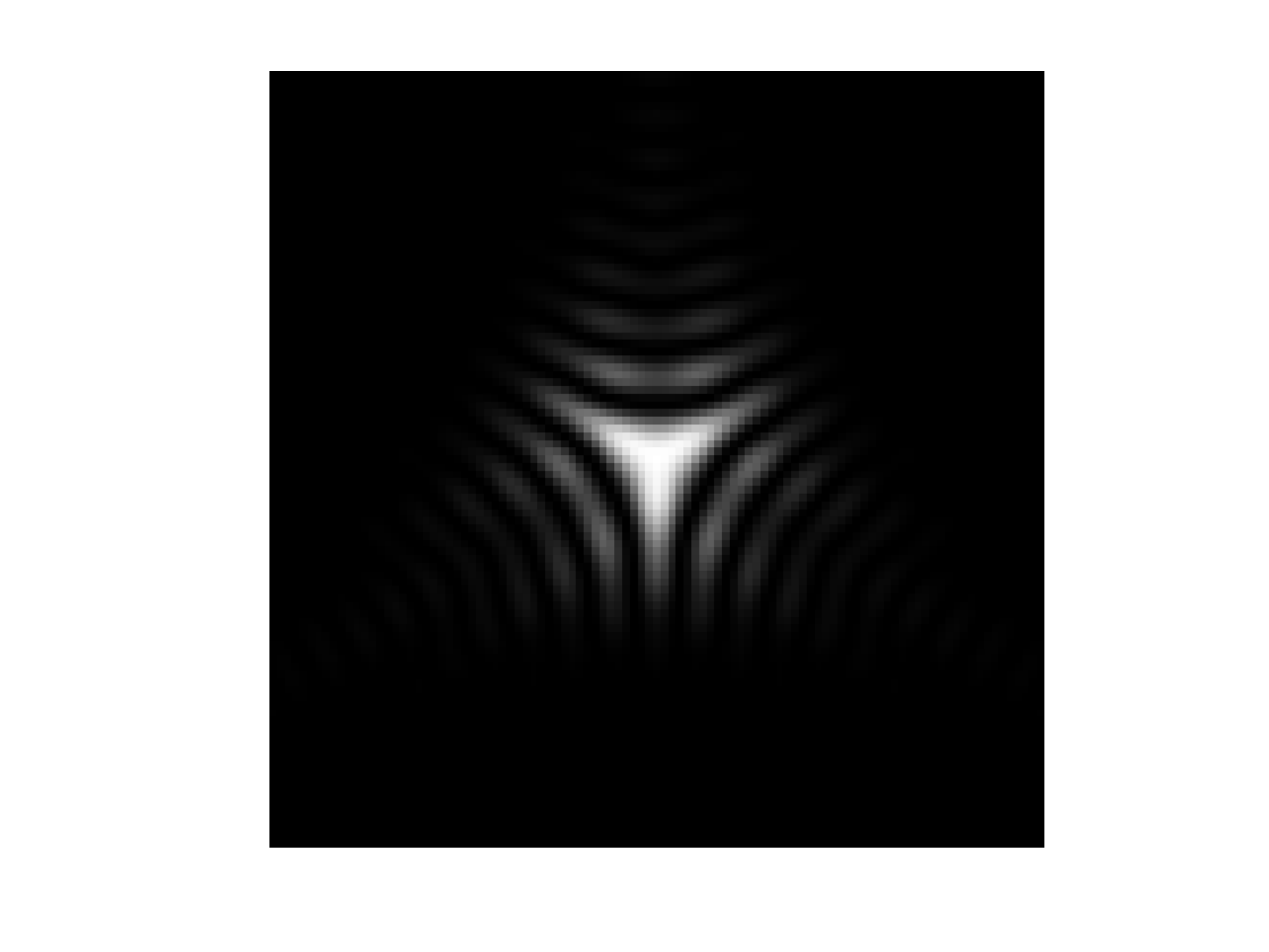}}
  		\label{fig:sig1C}
  	}
  	\end{minipage}\begin{minipage}[b]{0.5\linewidth}
  		\centering
  		\subfigure[]{
  		{\includegraphics[trim = 110 40 90 25, clip, width = 0.97\linewidth]{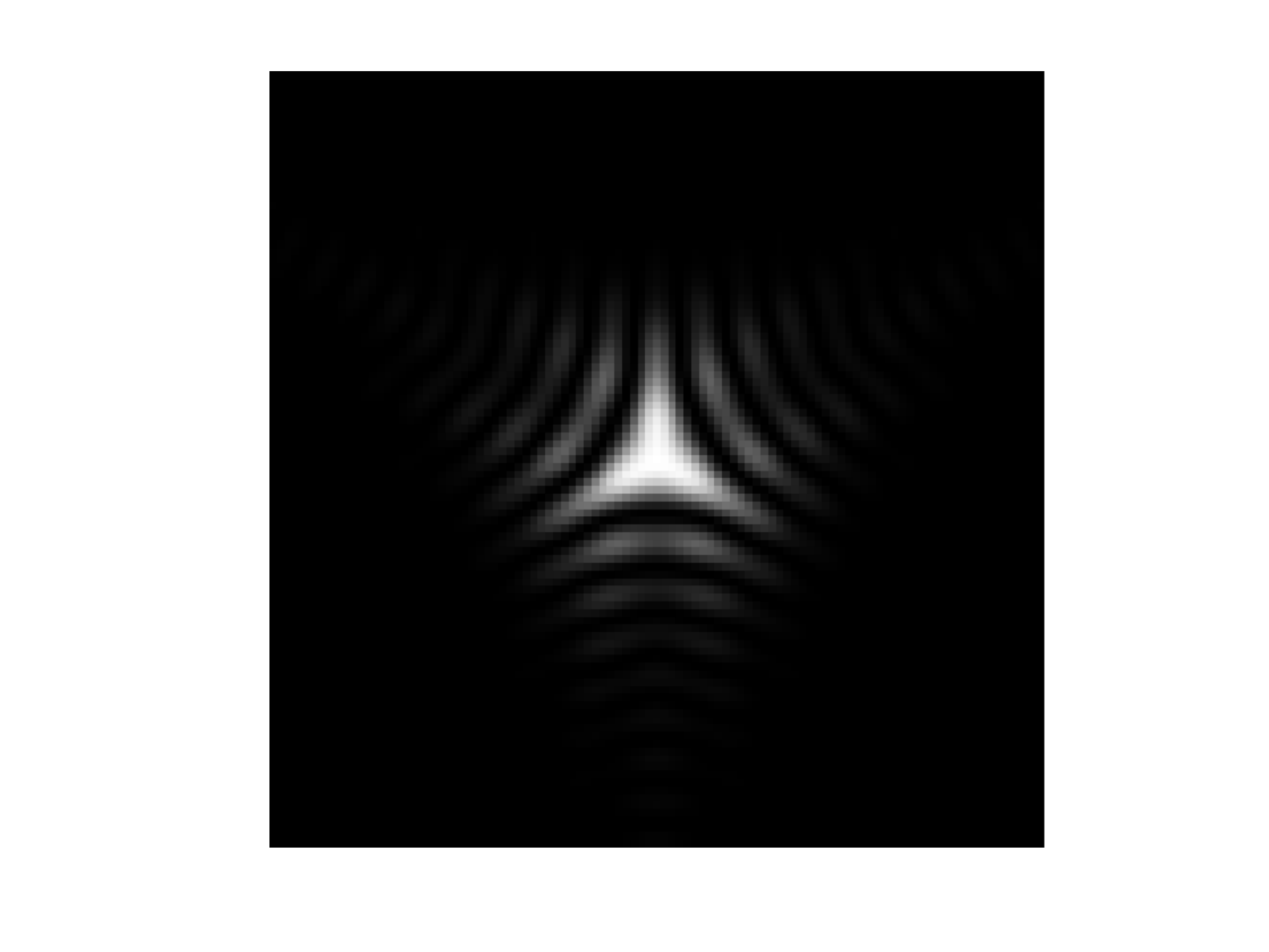}}
  		\label{fig:sig1D}
  	}
  	\end{minipage}
    \caption{The PSF intensity images of a point object with: (a) Zernike coefficient value $+5$ for $Z_{12}$ (b) Zernike coefficient value $-5$ for $Z_{12}$ (c) Zernike coefficient value $+5$ for $Z_{6}$ (d) Zernike coefficient value $-5$ for $Z_{6}$.}
    \label{fig:signed coeff}
\end{figure*}

As an example, consider Figure \ref{fig:signed coeff} which shows the PSF intensity images for a point object modeled using Zernike polynomials $Z_{12}$ and $Z_{6}$, respectively, with oppositely signed Zernike coefficients. Observe that for the angularly even polynomial $Z_{12}$ the intensity image is the same in \ref{fig:sig1A} and \ref{fig:sig1B} regardless of the sign of the Zernike coefficient. On the contrary, for the angularly odd polynomial $Z_{6}$, the intensity image is different in \ref{fig:sig1C} and \ref{fig:sig1D} as a result of the sign of the Zernike coefficient and thus does not result in an ambiguity.

\end{document}